\documentclass[fleqn,usenatbib]{mnras}

\usepackage{newtxtext,newtxmath}
\usepackage{xcolor} 

\usepackage[T1]{fontenc}

\DeclareRobustCommand{\VAN}[3]{#2}
\let\VANthebibliography\thebibliography
\def\thebibliography{\DeclareRobustCommand{\VAN}[3]{##3}\VANthebibliography}

\usepackage{graphicx}	% Including figure files
\usepackage{amsmath}	% Advanced maths commands
\usepackage{bbding}
\usepackage{hyperref}
\usepackage{scalerel,tikz}
\usetikzlibrary{svg.path}
\definecolor{orcidlogocol}{HTML}{A6CE39}
\tikzset{orcidlogo/.pic={
 \fill[orcidlogocol] svg{M256,128c0,70.7-57.3,128-128,128C57.3,256,0,198.7,0,128C0,57.3,57.3,0,128,0C198.7,0,256,57.3,256,128z};
 \fill[white] svg{M86.3,186.2H70.9V79.1h15.4v48.4V186.2z}
 svg{M108.9,79.1h41.6c39.6,0,57,28.3,57,53.6c0,27.5-21.5,53.6-56.8,53.6h-41.8V79.1z M124.3,172.4h24.5c34.9,0,42.9-26.5,42.9-39.7c0-21.5-13.7-39.7-43.7-39.7h-23.7V172.4z}
 svg{M88.7,56.8c0,5.5-4.5,10.1-10.1,10.1c-5.6,0-10.1-4.6-10.1-10.1c0-5.6,4.5-10.1,10.1-10.1C84.2,46.7,88.7,51.3,88.7,56.8z};
}}
\newcommand\orcidicon[1]{\href{https://orcid.org/#1}{\mbox{\scalerel*{
\begin{tikzpicture}[yscale=-1,transform shape]
\pic{orcidlogo};
\end{tikzpicture}
}{|}}}}

\title[Galaxy mass estimation using ML]{Galaxy stellar and total mass estimation using machine learning }

\author[Chu et al.]{
Jiani Chu\orcidicon{0009-0008-6513-0427}$^{1}$\thanks{E-mail: zjn20@mails.tsinghua.edu.cn},
Hongming Tang\orcidicon{0000-0002-7300-9239}$^{1}$\thanks{E-mail: hongmingt@tsinghua.edu.cn},
Dandan Xu$^{1}$, %$\thanks{E-mail: dandanxu@tsinghua.edu.cn},
Shengdong Lu\orcidicon{0000-0002-6726-9499}$^{2}$,
Richard Long\orcidicon{0000-0002-8559-0067}$^{1,3}$\\
$^{1}$Department of Astronomy, Tsinghua University, Beijing, 100084, China\\
$^{2}$Institute for Computational Cosmology, Department of Physics, Durham University, South Road, Durham, DH1 3LE, UK\\
$^{3}$Jodrell Bank Centre for Astrophysics, Department of Physics and Astronomy, The University of Manchester, Oxford Road, Manchester M13 9PL, UK\\
}

\date{Accepted XXX. Received YYY; in original form ZZZ}

\pubyear{2023}

\begin{document}
\label{firstpage}
\pagerange{\pageref{firstpage}--\pageref{lastpage}}
\maketitle

\begin{abstract}

Conventional galaxy mass estimation methods suffer from model assumptions and degeneracies. Machine learning, which reduces the reliance on such assumptions, can be used to determine how well present-day observations can yield predictions for the distributions of stellar and dark matter. In this work, we use a general sample of galaxies from the TNG100 simulation to investigate the ability of multi-branch convolutional neural network (CNN) based machine learning methods to predict the central (i.e., within $1-2$ effective radii) stellar and total masses, and the stellar mass-to-light ratio $M_*/L$. These models take galaxy images and spatially-resolved mean velocity and velocity dispersion maps as inputs. Such CNN-based models can in general break the degeneracy between baryonic and dark matter in the sense that the model can make reliable predictions on the individual contributions of each component. For example, with $r$-band images and two galaxy kinematic maps as inputs, our model predicting $M_*/L$ has a prediction uncertainty of 0.04 dex. Moreover, to investigate which (global) features significantly contribute to the correct predictions of the properties above, we utilize a gradient boosting machine. We find that galaxy luminosity dominates the prediction of all masses in the central regions, with stellar velocity dispersion coming next. We also investigate the main contributing features when predicting stellar and dark matter mass fractions ($f_*$, $f_{\rm DM}$) and the dark matter mass $M_{\rm DM}$, and discuss the underlying astrophysics.

\end{abstract}


%%%%%%%%%%%%%%%%%%%%%%%%%%%%%%%%%%%%%%%%%%%%%%%%%%

%%%%%%%%%%%%%%%%% BODY OF PAPER %%%%%%%%%%%%%%%%%%

\section{Introduction}%
\label{sec:introduction}

Achieving a full understanding of galaxy evolution requires accurate measurements of the ``unseen'' matter. This is why, among the many areas in astrophysical measurements and modelling, galaxy dynamics and gravitational lensing play unique roles, as they provide meaningful constraints on dark matter. On the other side of these measurements lies the distribution of the stellar component. Properties, such as stellar mass-to-light ratio $M_*/L$ and Initial Mass Function (IMF), are essential properties to solving the galaxy evolution puzzle. 
Therefore a fundamental task in this regard comes down to accurate determinations of the different contributions of dark matter and baryons -- a central goal of galaxy dynamics and gravitational lensing studies.   

In this regard, recent integral field spectroscopic observations have provided good datasets to study the dynamical properties of galaxies for a large sample of galaxies across a wide range of Hubble types, both in the nearby Universe, e.g., those from the $\rm ATLAS^{3D}$ (\citealt{Cappellari_2011MNRAS_ATLAS_I}), MaNGA (\citealt{Bundy_2015ApJ_MaNGA_Overview}), 
and SAMI (\citealt{Fogarty_2014MNRAS_SAMI}) surveys, and at high redshifts, e.g., the KMOS Galaxy Evolution Survey (KGES, \citealt{Turner17_KMOS-1}). Historically, people have developed various dynamical modelling methods (e.g., \citealt{Jeans_1922MNRAS, Schwarzschild_1979ApJ_SchwMod, Syer_1996MNRAS_M2M}), which typically combine a single-band image of a galaxy with stellar IFU-kinematic maps to constrain matter distributions across the galaxy. Specifically, these methods routinely split the total matter distribution into dark matter and baryonic matter components with each described by a pre-specified density profile. Modelling is then concerned with disentangling the individual contributions from both components, with specific objectives often set on estimating the central dark matter fraction or the IMF of the galaxy. The latter objective is often accomplished in combination with Stellar Population Synthesis (SPS) utilizing multi-band photometry or spectroscopic data. Such models have been widely applied to existing IFU galaxy surveys, for example, \citet{ZhuLing18_Nature} adopted the orbit-based Schwarzschild modelling techniques (\citealt{Schwarzschild_1979ApJ_SchwMod}) to the stellar kinematic data of 300 galaxies in the CALIFA survey (\citealt{2012A&A...538A...8S}). 
\citet{2012MNRAS.421.2580L} applied the made-to-measure (M2M, \citealt{M2Mmethod}) dynamical modelling technique to the SAURON (Spectrographic Areal Unit for Research on Optical Nebulae; \citealt{Cappellari2016SAURON}) galaxies. 
%\sdlu{\em{References of SAURON project would be better than the footnote.}}
Recently, \citet{Zhu2023_MaNGADynPopI} applied the Jeans Anisotropic Modelling (JAM, \citealt{Cappellari_2008MNRAS_JAM,Cappellari_2020MNRAS_JAM}) to the complete MaNGA sample (over 10K nearby galaxies with different morphologies) and obtained their `quality-assessed' dynamical properties.

One must be aware that even for the most sophisticated modelling techniques, there exist assumptions and approximations, which may lead to biased estimates on different levels. In terms of galaxy dynamical modelling, 
as full six-dimensional phase space data can not be obtained with current observational capabilities, dynamical properties are only inferred from projected light and kinematics. As a result, faithful estimation of dynamical, spatial, and orbital properties is not achievable over a wide range of cases. For example, Schwarzschild-based dynamical studies show that edge-on projections are preferred as more kinematic information is available (e.g., \citealt{ZhuLing18_CALIFA, ZhuLing2020_MethodValidation}), and inclined galaxies are more difficult to model accurately. In all cases, one must first make assumptions about a galaxy's matter geometry (e.g., spherical, elliptical, axisymmetric, or triaxial etc). In the case of Jeans-based methods, one must also make assumptions on the shape of velocity dispersion ellipsoid. The inferred results may then differ under different model assumptions. For example, as shown in Figure 10 
of \citet{Xu2017_IllustrisETG}, radial isotropy tends to underestimate the logarithmic slope of a galaxy's total density distribution in the inner region, while tangential isotropy tends to give overestimated results, when testing with simulated galaxies. 
{Note that there are limitations introduced by Jeans modelling and by the weighting schemes used in the Schwarzschild and M2M methods.
Jeans modelling may produce non-physical models (\citealt{Cappellari_2008MNRAS_JAM}, section 3.1.1) while the weighting schemes determine weights that
are numerically satisfactory in an optimisation context but may not be so astrophysically.

In addition, one must also take into account that information about a galaxy's light distribution does NOT directly equate to that of the baryonic mass distribution, unless it is known exactly how one is related to the other. An average converting factor, the baryonic mass-to-light ratio, could be obtained from the stellar mass-to-light ratio $M_{*}/L$ with a correction for gas, e.g., via an assumed stellar mass-to-gas mass scaling relation. The stellar mass-to-light ratio $M_*/L$, therefore, is a fundamental attribute, upon which estimates of many other properties may depend, e.g., the dark matter fraction. As we know, $M_*/L$ is neither a fixed value across a single galaxy nor some universal distribution across the galaxy population (e.g., \citealt{van_Dokkum_2010Natur_LowMassStarPopulation, van_Dokkum_2017ApJ_ETG_IMF_AbsorptionLine_III, LiHongyu_2017ApJ_MaNGA_IMF_variation,
Oldham_2018MNRAS_M87_IMF,  
ZhouShuang_2019MNRAS_MaNGA_IMF_Variation,Lu2023_MaNGADynPopII}). It depends on many galaxy properties such as IMF, star formation history, and so on, which may vary spatially, evolve with time, and depend on galaxy types. 
From observations, estimating the 3D stellar mass distribution from the light distribution may be attempted
using stellar population synthesis with 2D spatially resolved, spectroscopic data such that IMF-sensitive
absorption line features can help indicate the underlying stellar population and help constrain the
stellar mass-to-light ratio profile and thus approximate the stellar mass distribution (e.g, \citealt{van_Dokkum_2010Natur_LowMassStarPopulation, Spiniello_2014MNRAS_ETG_IMF, Parikh_2018MNRAS_MaNGA_IMF_spatial, Bernardi_2023MNRAS_MaNGA_IMF_systematic}).
However, such analyses are not straightforward to achieve at low cost for the majority of galaxies at all redshifts
and have their own degeneracies and shortcomings, for example, arising from lack of 3D observational data.

In conventional dynamical modelling, people commonly approximate $M_*/L$ with a constant value across a galaxy (e.g., \citealt{Cappellari_2011MNRAS_ATLAS_I, Zhu2023_MaNGADynPopI}), though not always (e.g.,\citealt{Oldham_2018_radialvar_IMF}).  
Unsurprisingly, observations indicate that constant $M_*/L$ may not be a universally good assumption across the galaxy population (e.g., \citealt{Tortora_2011MNRAS_M-L_grad, Garcia-Benito_2019A&A_M-L_sptaial_CALIFA, GeJunqiang_2021MNRAS_MaNGA_M-L_color,Lu2023_MaNGADynPopII}). Some studies, in particular those which combine stellar kinematics with gravitational lensing measurements for galaxies at higher redshifts, have adopted a power-law model to describe an imposed radial dependence of $M_*/L$, and attempted to estimate the power-law index from the observed data (e.g., \citealt{Sonnenfeld_2018MNRAS_SL_WL, Oldham_2018MNRAS_SL, Shajib_2021MNRAS_NFW_DMHalo}). 
The situation becomes even more complicated when the choice of different light and dark matter density models is taken into account. As neither of these $M_*/L$ assumptions (neither constant nor power-law) represents the true distribution, model fitting  under such assumptions lacks the power to select the right density model that is truly responsible for generating the observational data. This leads to model degeneracies being artificially broken, and causes the results to sensitively depend on the specific choice of light and/or dark matter density model, reaching biased estimates on either the stellar mass and thus the derived IMF or the central dark matter fraction. 
What is more, the absence of 6D data means we know nothing about the precise 3D spatial distribution of matter in a galaxy. This bias has been indeed manifested when tested against simulation galaxies for which ground truth values are known (e.g., \citealt{LiHongyu_2016MNRAS_IllustrisJAM}), or when tested against different model implementations on the same observational data (e.g., see Figure 12 of \citealt{Zhu2023_MaNGADynPopI}), and in some cases even by contradicting results obtained for similar galaxy populations (under the same $M_*/L$ assumption), but through different choices of the light model adopted (see \citealt{Sonnenfeld_2018MNRAS_SL_WL, Shajib_2021MNRAS_NFW_DMHalo}).

Machine learning provides an alternative way to start tackling the galaxy evolution problem, 
and has the advantage of making estimates of galaxy properties while eliminating many of the previous modelling assumptions. In addition, it has also been widely used as a powerful tool to understand the significant physical properties that link to cosmic structure formation and galaxy evolution. More and more studies have taken such approaches, from simply making predictions to certain properties (e.g., for galaxies, by \citealt{2019A&A...622A.137B}; for galaxy clusters, by e.g.,\citealt{2019MNRAS.484.1526A,2019ApJ...887...25H}), to inferring cosmological models and parameters (e.g., \citealt{2020PhRvD.101l3525A}), from emulating cosmic structure growth (e.g., \citealt{2019PNAS..11613825H, 2019ApJ...881...74M, 2021arXiv210508081T, 2021MNRAS.507.2510C}) to searching for physical connections between the predicted properties and input observational features (e.g., \citealt{dobbels2019, 2022MNRAS.515.2164L}). In a recent study by \citet{2023MNRAS.523.5408A}, galaxy populations from the TNG100\footnote{\url{www.tng-project.org}} (\citealt{Genel_2018MNRAS_TNG_SizeEvo, Nelson18_TNGcolor, Pillepich_et_al.(2018b), Springel_et_al.(2018), Marinacci18_TNGmagnetic, Naiman_et_al.(2018)}) and EAGLE simulations (\citealt{2015MNRAS.446..521S}) were used to calibrate a machine learning approach, which successfully predicted the fraction of accreted stars in galaxies from IFU-like observations. \citet{2023arXiv230710381G} trained a neural network as an emulator, massively speeding up likelihood evaluation for sophisticated and expensive dynamical modelling (around 200 times faster than similar emulations using JAM). In \citet{Hern_ndez_2023}, the stellar mass and Star-Formation Rate (SFR) 
of galaxies from the TNG300 simulation (\citealt{Nelson18_TNGcolor, Pillepich_et_al.(2018b), Springel_et_al.(2018), Marinacci18_TNGmagnetic, Naiman_et_al.(2018)}) were predicted using a neural network, which took as input 12 properties of galactic halos and their nearby environments. It was found that certain merger tree properties contribute significantly to the results from their machine learning model. \citet{Nicola2023RF_Mtotal} used a Random Forest (RF) based machine learning algorithm on TNG100 to predict the total and dark matter masses of galaxies with several simple observables as input, and then tested their approach on real galaxies. The results of their RF-based algorithm are consistent with the dynamical masses of real samples, and show the great potential of machine learning to make realistic estimates of galaxy masses. 
The \citet{Euclid2023preparationMLHimage} explored the potential of machine learning to estimate galaxy properties such as redshift, stellar mass, and SFR with data from the Euclid \citep{2011Euclid} and Rubin/LSST \citep{2008LSST} surveys.  They found that their models performed better in accuracy than spectral energy distribution modelling when predicting these properties.

Our goal in this study is not to develop any specific models to be applied to observational data, but to address the particular question as to whether existing or future observations might provide us with sufficient information for us to correctly disentangle individual mass contributions from baryons and dark matter. If yes, what are the reasons for such achievement; if no, again, what are the reasons? To do so, we take galaxies from the state-of-the-art cosmological hydrodynamical simulation -- the TNG100 simulation, and make mock observations of photometric images and IFU-like velocity maps for these galaxies. We utilize a Convolutional Neural Network \citep[CNN;][]{fukushima1982,lecun1998,krizhevsky2012,he2016deep} model to predict the stellar mass $M_*$ and total mass $M_{\rm tot}$ of galaxies that are enclosed within one half-stellar-mass spherical radius $R_{\rm hsm}$, as well as the stellar mass-to-light ratio $M_*/L$. We note that the detailed stellar and total mass density distributions of TNG100 galaxies are not precisely consistent with those observed (e.g., \citealt{Romeo2020SHMR_simulation_obs, LuShengdong_2020MNRAS_TNG_FP}). Therefore, we only employ the ML methods investigated in this study purely on the simulation dataset as a proof-of-concept study. Implementation to real observational data will also require further investigations examining observational effects, selection rules and so on. 
%This is because a previous study by \citet{Romeo2020SHMR_simulation_obs} pointed out that galaxies' stellar-to-halo mass fraction in EAGLE and TNG100 is smaller than observed galaxies and has a smaller variance. 

Results from our GPU-based CNN models suggest that 
the machine learning approach, for our simulated galaxies, is able to untwine the individual contributions from both dark and baryonic matter, from input maps of a galaxy's surface brightness distribution and its first and second line-of-sight velocity moments. To reveal any key factors which lead the CNN model to be able to make successful predictions, we use summary statistics from the images and maps as input to a Gradient Boosting Decision Tree model \citep[GBDT;][]{friedman2001,ke2017} to predict the values of the same galaxy properties.

The structure of the rest of this paper is as follows. 
In Section~\ref{sec:samples}, we introduce the IllustrisTNG galaxy sample that we use for this study, and how we build the datasets suitable for machine learning model training, testing, and interpretation. In Section~\ref{sec:methodology}, we give detailed descriptions of the two machine learning methods (CNN and GBDT) that we use. We show our results in Section~\ref{sec:CNN_results} (for CNN) and \ref{sec:GBDT_results} (for GBDT). Finally, we present discussions of our results and our overall conclusions in Section~\ref{sec:conclusion}.

\section{Sample Selection and Dataset Foundation}
\label{sec:samples}

As mentioned in Section~\ref{sec:introduction}, this work aims to test the feasibility and fidelity of two supervised machine learning methods (CNN, GBDT) in predicting the stellar, dark-matter, and total masses of galaxies, and the galaxies' mass-to-light ratios. 
In order to do this, we require a galaxy sample where these properties are known.  
Given that observations of real galaxies may suffer from various systematic biases, 
we choose to extract observationally equivalent data values from realistic galaxy simulations. In Section\,\ref{sec:sample selection}, we introduce the simulation-based galaxy sample used in this study. In Sections\,\ref{sec:CNN_input} and \ref{sec:GBDT_input}, we describe how we organize the necessary input data and targets suitable for CNN and GBDT modelling.

\subsection{Sample selection}
\label{sec:sample selection}
Our galaxy sample comes from the TNG100 simulation \citep{Genel_2018MNRAS_TNG_SizeEvo, Nelson18_TNGcolor, Pillepich_et_al.(2018b), Springel_et_al.(2018), Marinacci18_TNGmagnetic, Naiman_et_al.(2018)}, which is a set of magnetohydrodynamical (MHD) cosmological simulations of galaxy formation and evolution, using the {\sc arepo} software \citep{2010MNRAS.401..791S}. The simulation has been shown to broadly agree with many observed galaxy properties and general scaling relations, including the bimodal colour distribution \citep{Nelson18_TNGcolor}, the mass-size relation \citep{Genel_2018MNRAS_TNG_SizeEvo}, the galaxy mass density profiles \citep{WangYunchong_2019MNRAS_TNG_ETG_II, WangYunchong_2020MNRAS_TNG_ETG_I}, the fundamental plane relation \citep{LuShengdong_2020MNRAS_TNG_FP}, the dark matter fractions \citep{Lovell_2018MNRAS_TNG_fDM}, as well as the stellar orbit compositions \citep{XuDandan_2019MNRAS_TNG}. Specifically, the simulation has a box volume of $(\rm 110.7\,Mpc)^3$, a mass resolution of $\rm 1.4\times 10^6\,M_{\sun}$ and $\rm 7.5\times 10^6\, M_{\sun}$ for baryons and dark matter, respectively, and a force softening length of $\rm 0.5\,h^{-1}$kpc. 
The {\sc subfind} algorithm \citep{2001MNRAS.328..726S, 2009MNRAS.399..497D} is used to identify galaxies and their dark matter halos. General galaxy properties are available from \citet{2019ComAC...6....2N}.

\begin{figure*}
\centering
\includegraphics[width=0.24\textwidth]{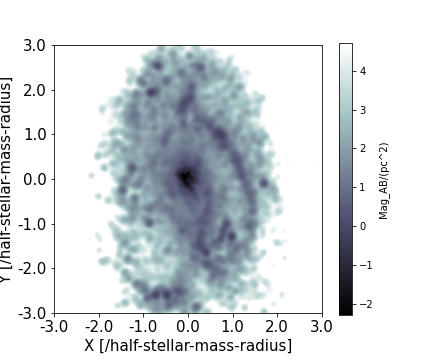}
\includegraphics[width=0.24\textwidth]{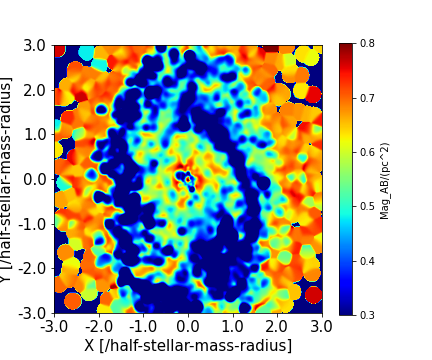}
\includegraphics[width=0.24\textwidth]{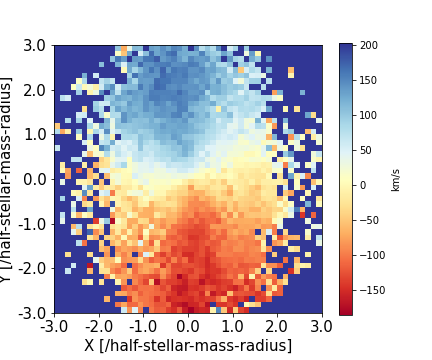}
\includegraphics[width=0.24\textwidth]{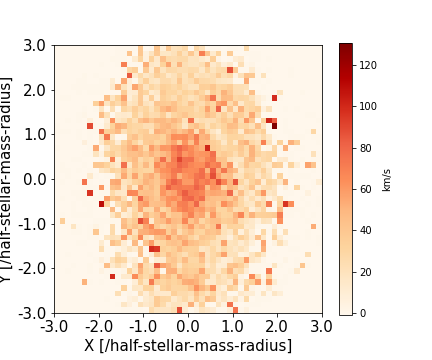}

\centering
\caption{From left to right, the four panels present the $r$-band image,  the $g$-$r$ colour map, the line-of-sight mean velocity map, and velocity dispersion map of an example galaxy (subhalo-ID: 501761). 
All images and velocity maps are produced in the range of $\pm 3R_{\rm hsm}$ from the galaxy centre. 
They are the basic input data set fed to our CNN-based model as Fig.\,\ref{fig:ResNet} shows. }
\label{fig:samples}
\end{figure*}

We take all galaxies 
at redshift $z \rm =0$ which have stellar mass within 30 kpc 
greater than $5\times 10^9\,\rm M_{\odot}$, and total subhalo mass (as calculated by {\sc subfind}) %denoted as $M_{\rm 200}$) 
less than $10^{14}\,\rm M_{\odot}$ 
The lower limit is to guarantee sufficient resolution, and the upper limit is to exclude systems in galaxy cluster environments, which are beyond the galaxy-mass range investigated in this study. To mimic the random orientation effect of observed galaxies, we just use the orientation of galaxies in the simulation. We project individual galaxies along the three principal axes of the simulation box, and take each projection as an independent galaxy in our sample. This projection operation enlarges our sample size, balancing dataset complexity (\citealt{2021data_complexity_inblanced_dataset}) and model complexity (\citealt{modelcomplexitysurvey}). 
The final $z=0$ dataset contains a total of 28110 galaxies (i.e., 9370 unique galaxies with three different projections of each).

\subsection{Data input, target generation and pre-processing for CNN}
\label{sec:CNN_input}
For CNN modelling, our input data for a galaxy comprises its $r$-band image, its $g$-$r$ colour map, the spatial distributions of stellar line-of-sight (along the direction of projection, i.e., the simulation axes) mean velocity and velocity dispersion for a given projection as they all contribute to the estimation of galaxy masses and stellar mass-to-light ratio \citep{binney+2008galaxy_dynamics,dobbels2019}. 
} 
Note that the first and second moments of line-of-sight velocities are directly calculated from stellar particles in the simulated galaxies. In this sense, these quantities do not have the same kinds of measurement errors as those derived from spectral line fittings. For simplicity, we do not consider the third and fourth velocity moments ($h_3$ and $h_4$). 
We note that all the data used in this work were extracted using pipelines developed for various previous studies by the authors (\citealt{Xu2017_IllustrisETG,
XuDandan_2019MNRAS_TNG, LuShengdong_2020MNRAS_TNG_FP, 2021MNRAS.503..726L, 2022MNRAS.509.5062L}). In particular, the spatial range and resolution of the kinematic maps of the simulated galaxies generally resemble typical SDSS and MaNGA-IFU observations. For MaNGA galaxies, the IFU observations for the stellar kinematic maps typically have a spatial coverage within $1.5-2.5$ effective radii from the galaxy centre \citep{2015ApJ...798....7B}. 
For all images and maps of the simulated galaxies, the spatial range was set to be within $\pm 3R_{\rm hsm}$ from the galaxy centre, where $R_{\rm hsm}$ is the 3D half-stellar-mass radius of the galaxy, roughly equivalent to the effective radius for an observed galaxy. Here below, we give a brief recapitulation of the techniques used to extract the data for the simulated galaxies.

For the $g$- and $r$-band images of the simulated galaxies, the luminosities of the stellar particles were processed for dust attenuation effects. This was carried out through a simple semi-analytical approach (see \citet{Xu2017_IllustrisETG} for details). 
Specifically, the $r$-band images and colour maps were produced in a mesh of 300\,$\times$\,300, corresponding to 0.02 $R_{\rm hsm}$ per pixel. This high spatial resolution allows the adoption of a cubic spline kernel as used in the Smoothed Particle Hydrodynamics \citep[SPH;][]{monaghan1992sph,1999A&A...347..769H}. The dust-attenuated luminosities are then assigned and smoothed via the SPH scheme into mesh pixels, with a smoothing length that encloses the nearest 32 neighbouring stellar particles. 
For our CNN resolution tests, the original SPH-smoothed images are re-binned to resolutions of 150\,$\times$\,150 and 60\,$\times$\,60. The last setting matches the median resolution for the SDSS galaxy survey.

For the kinematic maps, instead of adopting a Voronoi binning scheme,  for simplicity, we directly projected stellar particles on to a mesh using the Near-Grid-Point (NGP) scheme. 
These velocity maps have dimensions of 48\,$\times$\,48 pixels, corresponding to a spatial resolution of 8 pixel per $R_{\rm hsm}$, which corresponds to resolving a typical MaNGA galaxy at redshift $z=0.05$, and is slightly below the median value of 14 pixel per $R_{\rm hsm}$ for the entire MaNGA galaxy sample (derived from \citealt{manga_overview}). 
Fig.\,\ref{fig:samples} shows the above-mentioned images and maps for one example galaxy in our data set. Note that we do not add any observational errors to the generated images and maps.

For the training process on our CNN network, we must provide target galaxy data values for the properties we wish our network model to predict. In our case, we provide the central stellar mass ($M_*$), total mass ($M_{\rm tot}$), and mass-to-light ratio ($M_*/L \equiv M^{2D}_*/L$, where $L$ is the $r$-band dust-attenuated luminosity). For modelling real galaxies, estimating these values has always been the goal for conventional studies in stellar population synthesis, stellar kinematics, and gravitational lensing. In particular, the stellar mass (or the stellar mass-to-light ratio $M_*/L$) and total mass are the two most commonly derived basic quantities. Once they are obtained, the dark matter fraction $f_{\rm dm}$ can in principle be further derived. Here the mass values are determined using particles of the corresponding type, located within a 3D sphere of radius $R_{\rm hsm}$ from the galaxy centre. The $M_*/L$ value is a projected quantity and is calculated using the $r$-band luminosity and stellar mass of stellar particles projected within a radius of $R_{\rm hsm}$ for a given line-of-sight.  

Prior to using the data in CNN modelling, the data are cleaned (for example, to ensure it does not contain any spurious values) and formatted according to the requirements of the modelling software being used (PyTorch in our case). 
To ensure auto-diff (Automatic Differentiation) functions properly, we perform a normalization operation, which is to linearly scale all pixels of images and maps to ensure the numerical value of the pixels is between 0 and 1. 
In our models, we split our galaxy samples into 3 parts: a training set (16000 samples), a validation set (4000 samples), and a test set (8110 samples). Operationally, it is convenient, for comparison purposes, to ensure that the sets always contain the same galaxies.  This is achieved by setting a random seed to the same value in all modelling runs.

\subsection{Data input and target generation for GBDT}
\label{sec:GBDT_input}
For GBDT modelling \citep{friedman2001}, the model inputs are a number of summary statistics extracted from the particles of the simulated galaxies. We use the following quantities: the $r$-band absolute magnitude $M_r$ and $g$-$r$ colour of a galaxy, the star-formation rate, the S{\'e}rsic index $n_{\rm Ser}$, the stellar axis ratio $c/a$, the velocity dispersion $\sigma_{\rm v}$, the dimensionless spin attribute $\lambda$ (quantifying the degree of stellar rotation, see \citealt{2007MNRAS.379..401E} for a detailed definition), the kinetic bulge-to-total ratio $B/T$, and the cold and hot orbital fractions $f_{\rm cold}$ and $f_{\rm hot}$. Notice that orbit fractions can not be directly obtained without dynamical modelling. and are only used in GBDT tasks. Detailed descriptions of the quantities can be found in Table \ref{tab:GBDT_input}.
%$\lambda=\frac{<R\lvert V\rvert>}{<R\sqrt{V^2+\sigma^2}>}$, 

Using the above-mentioned quantities, we make predictions on the stellar mass $M_*$, dark matter mass $M_{\rm DM}$, and total mass $M_{\rm tot}$ of our galaxies, as well as on the mass-to-light ratio $M_*/L$ and dark matter fraction $f_{\rm dm}$. For training purposes, all these quantities are determined within one $R_{\rm hsm}$.

We exclude galaxies whose S{\'e}rsic index $n_{\rm Ser}$ is larger than 100 and whose kinetic bulge-to-total ratio $B/T$ is larger than 1. Such galaxies only make $\sim 3\%$ of the total sample size.

\begin{table}
    \centering
    \caption{Feature inputs of our GBDT model}
    \label{tab:GBDT_input}
    \begin{tabular}{l|l}
    \hline input & description \\
    \hline $M_r$ & SDSS $r$-band absolute AB magnitude \\
    $g$-$r$ & SDSS $g$-$r$ colour \\
    SFR & star forming rate over the past 1 Gyr within \\ 
    & a projected radius of $2R_{\rm hsm}$ from the galaxy centre\\

    $n_{\rm Ser}$  & S{\'e}rsic index from a S{\'e}rsic profile (\citealt{Sersic_1963BAAA}) fitting \\
    & to the light distribution within a projected radius of $5R_{\rm hsm}$ \\
    $c/a$ & shortest-to-longest axis ratio $c/a$ of the stellar mass distribution, \\
    & calculated using the inertial tensor method (\citealt{2006MNRAS.367.1781A}), \\
    & through an iterative approach started within a 3d radius of $3R_{\rm hsm}$ \\
    & (see \citep{2007MNRAS.379..401E} for detailed definition) \\
    $\sigma_{\rm v}$ & the the $r$-band luminosity weighted line-of-sight \\ 
    & velocity dispersion within a projected radius of $2R_{\rm hsm}$  \\
    $\lambda$ & dimensionless spin calculated within projected $2R_{\rm hsm}$ \\ 
    & (see \citet{2007MNRAS.379..401E} for detailed definition) \\
    $B/T$ & a kinetic bulge-to-total ratio within $2R_{\rm hsm}$, defined as \\ 
    & two times the mass fraction of stellar particles with negative \\ & circularity, i.e., $\epsilon<0$ (see \citet{XuDandan_2019MNRAS_TNG} for definition)  \\
    $f_{\rm cold}$ & mass fraction of stellar particles with cold orbits ($\epsilon>0.8$) \\ 
    & within a 3d radius of $2R_{\rm hsm}$ (see \citet{XuDandan_2019MNRAS_TNG} for definition) \\
    $f_{\rm hot}$ & mass fraction of stellar particles with hot orbits ($|\epsilon|<0.25$) \\
    & within a 3d radius of $2R_{\rm hsm}$ (see \citet{XuDandan_2019MNRAS_TNG} for definition) \\ \hline
    \end{tabular}
    
\end{table}

\section{Methodology}
\label{sec:methodology}
\begin{figure}
    \centering
    \includegraphics[width=1\columnwidth]{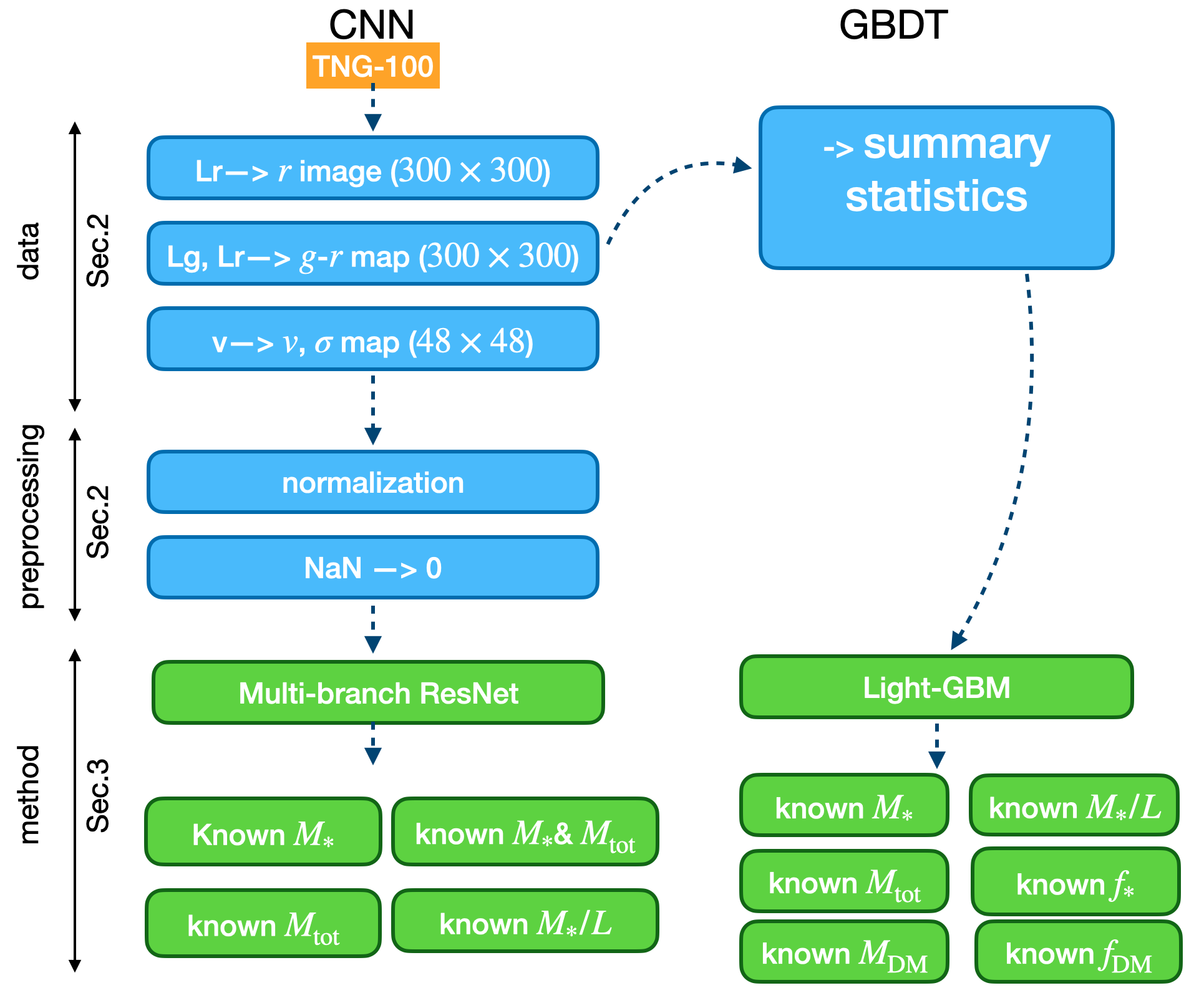}
    \caption{Workflow of this study: data, preprocessing, and model  tasks from top to bottom (left: CNN; right: GBDT). 
    %\sdlu{\em It seems that you did not mention ``nan to 0'' in you text. I think it would be important to present the reasons of your choice. And also, you should make sure that using 0 to replace NaN values would not influence your results.}
    }
    \label{fig:paper_structure}
\end{figure}

The workflow for our modelling is displayed in Fig.\,\ref{fig:paper_structure}. %\rjlnote{Text size is too small in figures 2 and 3.  Suggest replacing the black boxes in figure 3 by yellow / gold boxes.}
Section~\ref{sec:samples} covers the galaxy sample selection and data processing aspects of the workflow. In this section we introduce the model architecture and training setup of the two machine learning algorithms we use: CNN is described in Section~\ref{sec:CNN} and GBDT in Section~\ref{sec:GBDT}.

Given that we make use of both algorithms to predict numerical values, we utilize the same Mean Square Error (MSE) loss function (Equation\,\ref{MSE}) when training our models.
\begin{equation}
\label{MSE}
    {\rm MSE}=\frac{1}{N}\sum_i^N (y_i^{\rm pred}-y_i^{\rm true})^2,
\end{equation}
where $y_i^{\rm pred}$ and $y_i^{\rm true}$ are the predicted and the true values of the galaxy attributes, and $N$ is the sample size.

\subsection{Convolutional neural network: Multi-branch ResNet}
\label{sec:CNN}

CNNs are a type of Artificial Neural Network (ANN) making use of convolution filters (kernels) that enable them to capture 
features directly from input images, 
and are widely used in image processing (\citealt{arena2003image, han2020underwater, 2022MNRAS.516..264S, 2022SPIE12189E..1QN, 2020A&A...633A.148B}). In practice, a typical CNN would follow a top-down structure: some convolutional layers in linear sequence first extract features from model inputs, squeeze these features into a linear format, and forward them to sequenced fully-connected layers for further feature abstraction and then make a model prediction.

A common technique to improve a CNN model's performance is to increase the number of layers in the model. However, it was found that a model cannot always improve its performance by simply adding more network layers. Model classification accuracy may saturate and eventually suffer from rapid degradation \citep{he2016deep}. To resolve this so-called `degradation problem', \citealt{he2016deep} proposed Deep Residual Networks. This architecture introduced a `residual block' (see schematic in Figure\,2 of \citealt{he2016deep})—instead of optimizing the output of a stacked 2-layer block \textbf{F(x)}, a `residual block' asked the network to optimize the combination of block output \textbf{F(x)} and block input \textbf{x}, which gives \textbf{H(x) = F(x) + x}. This optimisation was believed to be easier to achieve \citep{he2016deep}. Such an innovation helped ResNet win the 2015 ImageNet Large Scale Visual Recognition Challenge \citep[ILSVRC15;][]{ILSVRC15}. ResNet has been used in earlier astronomical studies such as finding galaxy-galaxy strong lenses \citep{Lanusse2018}, and classifying galaxy clusters \citep{2020MNRAS.498.5620S}. With both model performance and computation power limitations in mind, we chose a modified version of ResNet-18 (ResNet-18 hereafter for simplicity; \citealt{he2016deep,2020MNRAS.498.5620S}) as our CNN backbone: the backbone contains one convolutional layer and 8 `residual blocks'.

\begin{figure}
\centering
\includegraphics[width=1\columnwidth]{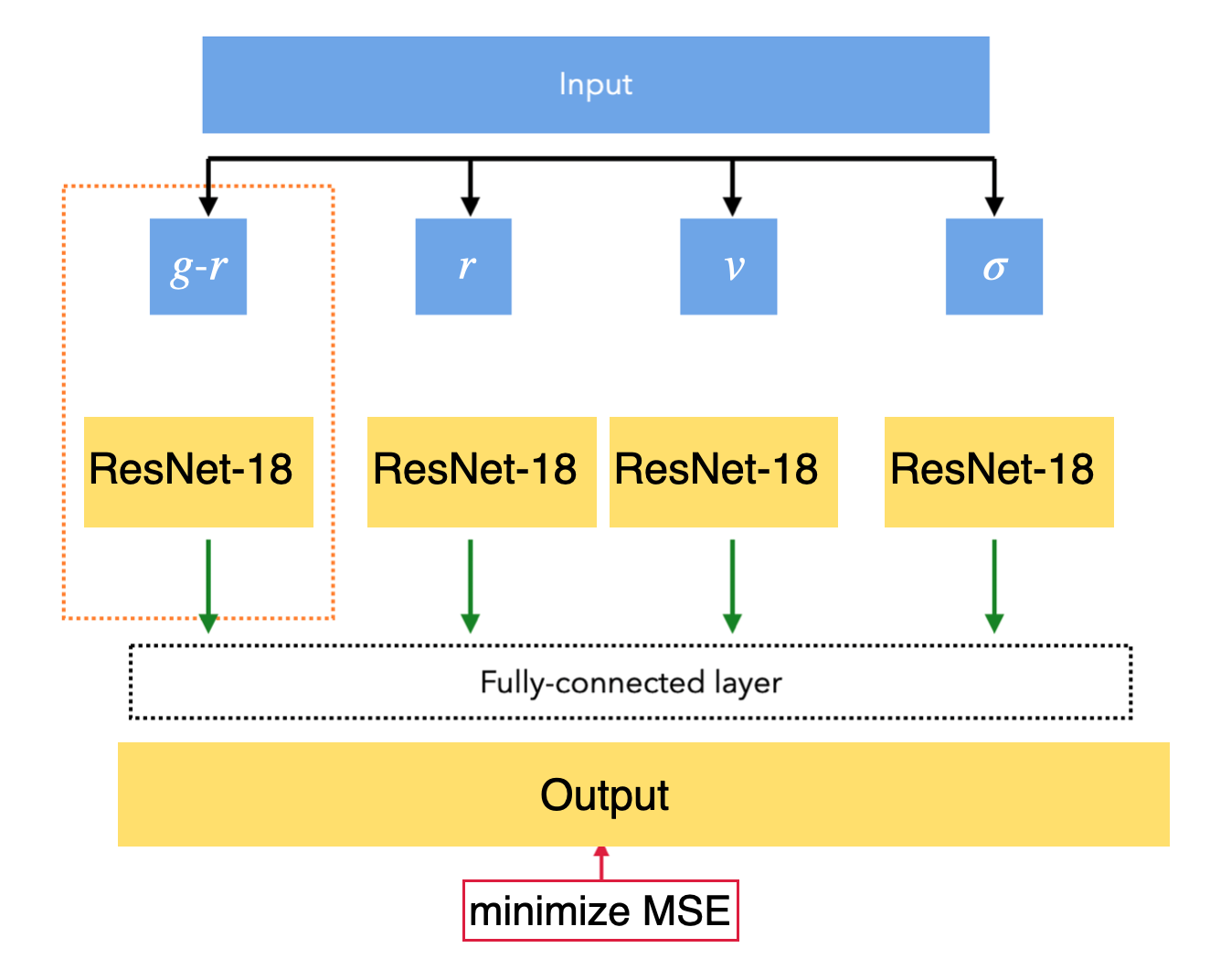} 
\caption{Structure of multi-branch ResNet in this work: $g$-$r$ map, $r$-band image, and velocity maps are processed by normal ResNet-18 backbone independently until the last fully-connected layer. At last, all intermediate outputs are combined to give predictions via a fully connected layer.}
\label{fig:ResNet}
\end{figure}

While a classic top-down CNN structure can extract features from a single image or map and make predictions, it cannot solve our requirement to use multiple images and maps with different mesh sizes. An approach to address such a requirement is to use a multi-branch neural network \citep{al2018learning}. Multi-branched network architectures allow one to simultaneously utilize multiple input data sets in one model for feature extraction and model prediction. 
Multi-branched networks have previously been used to identify lensed supernovae \citep{2022ApJ...927..109M} and giant radio galaxies \citep{tang2022}. \citet{tang2022} suggested implementing such architectures could boost model performance. In this work, we adopt a multi-branched network architecture for our CNN-based models. It can be seen from Fig,\,\ref{fig:ResNet} that our input images and maps are individually fed into ResNet-18 backbones. These backbones extract features from images/maps, and forward their outputs to a fully-connected layer to produce a model prediction.

When training models, we determine model hyperparameters through a manual selection process \citep{bergstra2012random}. This is achieved by training our model using the training dataset, and manually selecting model hyperparameters by looking at the model's behaviours when making predictions using samples in the validation set. The model's ability to generalize is evaluated using the test set of samples, as may be seen in Section\,\ref{sec:CNN_results}. 

After experimentation, we choose the Adam optimizer, which performs better than the Stochastic Gradient Descent (SGD) optimizer. To avoid over-fitting, we utilize a cosine learning rate scheduler \citep{Schaul2013} ($lr=0.0001 \cos (n/200)$, where $n$ is the training epoch number) instead of a constant learning rate. Table\,\ref{tab:ResNet_Hyperparam} lists all the hyperparameters of our multi-branch ResNet. A training batch size of 40 was chosen after consideration of the available GPU memory. As is common practice, we shuffle our sample input sequences before each model training epoch. 

The CNN-based model implementation in this work uses the PyTorch python library\footnote{\url{https://pytorch.org}}, where ResNet backbones (i.e. ResNet-18, ResNet-50) are inbuilt and ready to use.

\begin{table}
    \centering
    \caption{Hyperparameters of our multi-branch ResNet model}
    \label{tab:ResNet_Hyperparam}
    \begin{tabular}{c|c}
    \hline hyperparameter & setup \\
        \hline Batch size & 40  \\
         initial learning rate & 0.0001 \\
         learning rate step & 200 \\
         training epoch & 200 \\ \hline
    \end{tabular} 
    
\end{table}

\subsection{Gradient boosting machine: Light-GBM}
\label{sec:GBDT}

A Gradient Boosted Decision Tree (GBDT) is a machine learning method with an efficient catalogue data processing capability and high model interpretability. It builds a series of decision trees to address either classification or regression problems, and has been used in recent astronomical studies \citep[e.g.,][]{coronado2022gbdt,sahakyan2023gbdt}. GBDT attempts to build a `strong' model using multiple `weak' models (i.e., decision trees). 
The general procedure for GBDT training is as follows.
\begin{enumerate}
    \item The first tree is formed to fit the given training data and make predictions.
    \item A second tree is then formed to fit the residuals between the first tree's predicted values and the truth values.
    \item The next step is iterative where successive trees are trained to fit the residuals of the previous one.
    \item The model training process stops when some customized stopping criteria have been met.
\end{enumerate}

Although GBDT has been widely implemented, training a GBDT model can be time-consuming when the training data set has a large sample size or a considerable number of features. This is because the classic GBDT needs to scan all data points and estimate the information gain for every possible split point \citep{ke2017}. A representative approach that tackled this time issue is LightGBM \citep{ke2017}. This is an improved version of GBDT with two novel techniques introduced, Gradient-based One-Side Sampling (GOSS) and Exclusive Feature Bundling (EFB) (see \citet{ke2017} for full technical details). In short, GOSS allows LightGBM to estimate the information gain at its split points using those data samples with larger gradients, and EFB bundles mutually exclusive features to decrease the number of features. Compared with the original GBDT, LightGBM speeds up the model training process by up to 20 times and achieves similar accuracy \citet{ke2017}. Here we implement LightGBM to predict the attributes of our galaxies ($\rm M_*,M_{tot},M_{DM},M_*/L,f_*\,and\,f_{DM}$). We also investigate which summarized image or map features have contributed to the prediction of which attributes by performing feature importance evaluations of these LightGBM models (see Section~\ref{sec:GBDT_results}).

To train and test LightGBM-based models, we split our samples into two parts: the training set (18833 samples) and the test set (9277 samples). We do not consider validation sets or model cross-validation as these models are developed for proof-of-concept purposes. Rather than emphasising model accuracy and stability, we would like to investigate its ability to (a) break the degeneracy between baryonic and dark matter in galaxies, and (b) explore why different galaxies have different properties.

Instead of tuning hyperparameters, we conduct experiments with different input and output combinations, while maintaining the same set of hyperparameter values, to help us evaluate the behaviour of our GBDT models and to understand what summary statistics of images and maps have specific physical meanings. 
By comparing model behaviours trained with inputs of different ranges with fixed model training hyperparameters, we find that when using input features at 2$R_{\rm hsm}$ apertures instead of $1R_{\rm hsm}$, the resulting mass prediction uncertainty decreased by 9\%.  
This is why our inputs are mainly galaxy properties from $2 R_{\rm hsm}$ apertures, while our outputs are all at $1 R_{\rm hsm}$ apertures.

\begin{figure*}
\centering

\includegraphics[width=\textwidth]{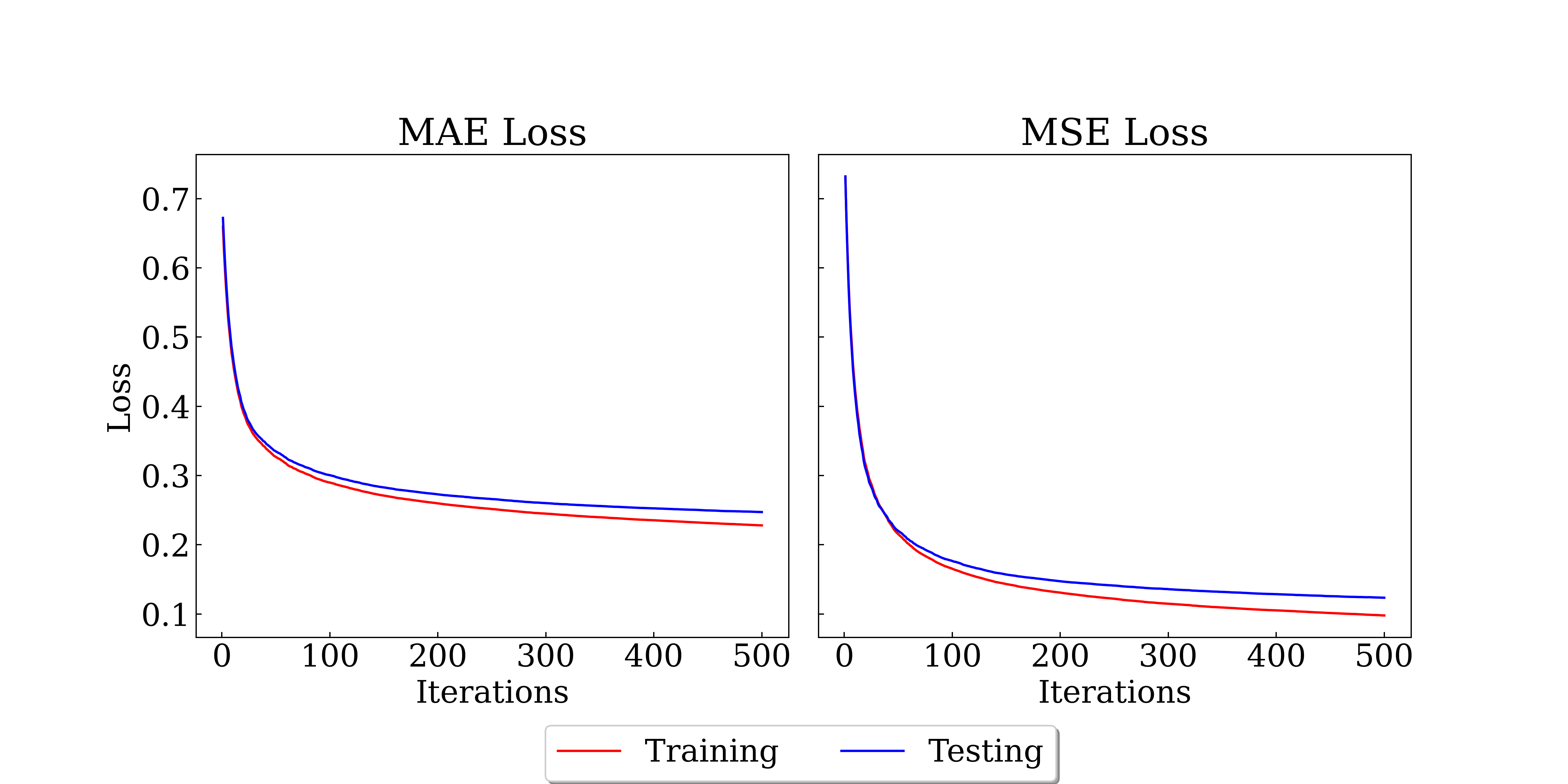}
\caption{Loss curves of LightGBM-based model as a function of training iterations with known $M_*/L$ (left: Mean Absolute Error (MAE) loss; right: Mean Square Error (MSE) loss). In each panel, the red and blue curves indicate the training and testing losses, respectively.}
\label{fig:lightgbm_mstarL_loss}
\end{figure*}

Table\,\ref{tab:GBDT_Hyperparam} lists the hyperparameters we use  (after experimentation) in our LightGBM models. We use 500 trees for training, with the maximum number of leaves in any node (the \textit{$ num\_leaves$} parameter) set to 5. We do not include further constraints on the minimum data sample number in any one leaf, the maximum depth of a tree (the typical depth in this work is 4), or apply other regularization methods as we do not observe any sign of model over-fitting. Notably, we have evaluated 
Mean Absolute Error (MAE) and MSE loss functions
to understand whether model robustness is affected by possible outliers in our data samples.  We find that training algorithms in this work behave well when looking at both loss curves (see Fig.\,\ref{fig:lightgbm_mstarL_loss} for an example).

\begin{table}
    \centering
     \caption{Hyper parameters of our GBDT model }
     \label{tab:GBDT_Hyperparam}
    \begin{tabular}{c|c}
    \hline Hyperparameter & setup \\
      \hline  num\_leaves & 5  \\
      num\_boost\_round & 500 \\
         objective &  regression \\
         min\_data\_in\_leaf & 20(default) \\
         max\_depth & no limit \\

         evaluation metrics & MAE and MSE \\\hline
    \end{tabular}
    
\end{table}

\section{CNN-based model results}
\label{sec:CNN_results}

In this section, we present the results from our CNN-based modelling. In all our models, we train the network using the multi-branch ResNet taking some combination of $r$-band images ($300\,\times\,300$ SPH-smoothed images within $\pm 3R_{\rm hsm}$) and the two velocity maps ($48\,\times\,48$ NGP-smoothed maps within $\pm 3R_{\rm hsm}$) for mean velocity and dispersion simultaneously as input, and a variety of galaxy mass and light quantities as targets.

\begin{figure}
\centering

\includegraphics[width=\columnwidth]{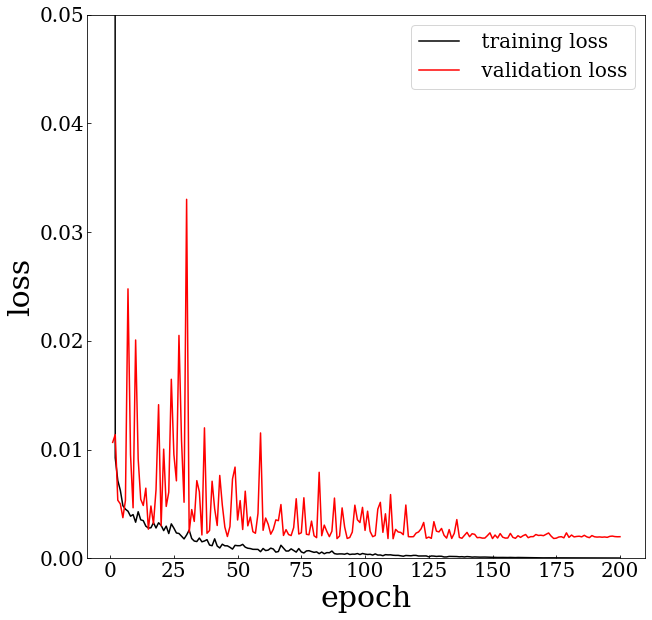}
\caption{MSE loss (see Section~\ref{sec:CNN_results_mstar}) curve of CNN-based model with known $M_*$ as function of training epoch. The black and red curves indicate the training and validation loss respectively.}
\label{fig:CNN_mstar_loss}
\end{figure}

As an example, Fig.\,\ref{fig:CNN_mstar_loss} shows the model training and validation loss curves using galaxy stellar mass $M_*$ as the target. 
As can be seen, while the model validation loss is high and oscillates considerably initially, it gradually decreases and becomes stable after $\sim$ 140 training epochs. 
We train our model for a total of 200 epochs and select the epoch where the validation loss minimizes our model. Doing so prevents a model from over-fitting the data. We save this model and evaluate its performance using the test data set. 
Instead of using MSE loss to evaluate our model performance, we use 1-$\sigma$ of log $y^{\rm pred}/y^{\rm true}$ to evaluate the uncertainty of the prediction, which can be understood as the scatter of the prediction.  
Here the uncertainty is intended to describe the performance of the model prediction, and is not related to observational errors. 

Table \ref{tab:CNN_performance_III} summarizes all the CNN models we run, and indicates the subsections in which the results are described.  We wish to point out that some of the tests are concerned with predicting attribute values within $2R_{\rm hsm}$ rather than just the $1R_{\rm hsm}$ we usually employ.

\begin{table*}
%    \captionof{Results of CNN-based model}
    \centering    
    \begin{tabular}{c|c|c|c|c}
    \hline  Section & Input & Training on / Output & Prediction & mean $\pm$ 1-$\sigma$ (dex) \\
      \hline\hline \ref{sec:CNN_results_mstar} &  $r$ image & $M_*$ & $M_*$  &0.00 $\pm$ 0.06  \\
      \hline \ref{sec:CNN_results_mstar} & $r$ image & $M_*^{\rm 2D}$ & $M_*^{\rm 2D}$ &-0.01$\pm$ 0.07 \\
       \hline \ref{sec:CNN_results_mstar} & $g$, $r$ image & $M_*$ & $M_*$  &0.00 $\pm$ 0.05  \\
       \hline \ref{sec:CNN_results_mstar} & $g$, $r$ image & $M_*^{\rm 2D}$ & $M_*^{\rm 2D}$ & 0.00$\pm$ 0.06\\
      \hline \ref{sec:CNN_results_mstar} &  2 V maps & $M_*$ & $M_*$  &0.00 $\pm$ 0.05  \\
      \hline \ref{sec:CNN_results_mstar} &  2 V maps & $M_*^{\rm 2D}$ & $M_*^{\rm 2D}$ &0.00$\pm$ 0.05\\
      \hline  \ref{sec:CNN_results_mstar}& $r$ image + 2 V maps & $M_*$ & $M_*$  &0.00 $\pm$ 0.04  \\
      \hline  \ref{sec:CNN_results_mstar}& $r$ image + 2 V maps & $M_*^{\rm 2D}$ & $M_*^{\rm 2D}$ &0.00$\pm$ 0.05\\
      %\hline  \ref{sec:CNN_results_mstar}& $r$ image + 2 V maps & $M_*$ & $M_*/L$ &0.00 $\pm$ 0.04  \\
      \hline  \ref{sec:CNN_results_mstar}&   $r$ image + 2 V maps & $M_*(\leqslant 2R_{\rm hsm})$ & $M_*(\leqslant 2R_{\rm hsm})$  &0.00 $\pm$ 0.04  \\ 
      \hline  \ref{sec:CNN_results_mstar}&   $r$ image + 2 V maps & $M_*^{\rm 2D}(\leqslant 2R_{\rm hsm})$ & $M_*^{\rm 2D}(\leqslant 2R_{\rm hsm})$ &0.00 $\pm$ 0.04  \\
      \hline\hline  \ref{sec:CNN_results_mtot} & $r$ image & $M_{\rm tot}$ & $M_{\rm tot}$  &0.00 $\pm$ 0.07  \\
      \hline \ref{sec:CNN_results_mtot} &  $g$, $r$ image & $M_{\rm tot}$ & $M_{\rm tot}$  &0.00 $\pm$ 0.07  \\
       \hline \ref{sec:CNN_results_mtot} &  2 V maps & $M_{\rm tot}$ & $M_{\rm tot}$  &0.00 $\pm$ 0.09  \\
      \hline  \ref{sec:CNN_results_mtot} & $r$ image + 2 V maps & $M_{\rm tot}$ &  $M_{\rm tot}$ & 0.00 $\pm$ 0.06  \\
      \hline \ref{sec:CNN_results_mtot} & $r$ image + 2 V maps & $M_{\rm tot}(\leqslant 2R_{\rm hsm})$ & $M_{\rm tot}(\leqslant 2R_{\rm hsm})$  &0.00 $\pm$ 0.06  \\
      \hline\hline  \ref{sec:CNN_results_2prop}& $r$ image + 2 V maps & $M_*$, $M_{\rm tot}$& $M_*$  &0.01 $\pm$ 0.04  \\
      \hline  \ref{sec:CNN_results_2prop}& $r$ image + 2 V maps & $M_*$, $M_{\rm tot}$& $M_{\rm tot}$ & -0.01 $\pm$ 0.06  \\
      \hline  \ref{sec:CNN_results_2prop}& $r$ image + 2 V maps & $M_*$, $M_{\rm tot}$& $f_{\rm DM}$ &0.02 $\pm$ 0.05  \\
      \hline  \ref{sec:CNN_results_2prop}& r image + 2 V maps & $M_*$, $M_{\rm tot}$& $f_{\rm DM}$ of ETGs &-0.02 $\pm$ 0.04  \\
      \hline\hline  \ref{sec:CNN_results_m2l} & $r$ image & $M_*/L$ & $M_*/L$  &0.01 $\pm$ 0.10 \\
      \hline \ref{sec:CNN_results_m2l}& 2 V maps & $M_*/L$ & $M_*/L$  &0.01 $\pm$ 0.11  \\
      \hline \ref{sec:CNN_results_m2l}& $r$ image + 2 V maps & $M_*/L$& $M_*/L$ & 0.01 $\pm$ 0.07  \\
       %\hline \ref{sec:CNN_results_m2l}& $r$ image + 2 V maps & $M_*/L(\leqslant 2R_{\rm hsm})$ & $M_*/L(\leqslant 2R_{\rm hsm})$  &0.00 $\pm$ 0.07  \\
      \hline \ref{sec:CNN_results_m2l}& $g$, $r$ image & $M_*/L$ & $M_*/L$  &0.01 $\pm$ 0.05  \\
      \hline  \ref{sec:CNN_results_m2l}& $g$, $r$ image + 2 V maps & $M_*/L$ & $M_*/L$ &0.00 $\pm$ 0.04  \\
      \hline\hline
    \end{tabular}
    \caption{Results of CNN-based models, where multi-branch ResNet takes $r$-band images and two velocity maps simultaneously as input (see Section\,\ref{sec:CNN_results} for details). The fifth column indicates the mean and standard deviation of the logarithmic ratio between the predicted and the true values for quantities given in the fourth column. All properties are evaluated within a radius of $R_{\rm hsm}$ from the galaxy centre, except for those explicitly specified with parentheses.  }
    \label{tab:CNN_performance_III}
\end{table*}

\subsection{Training CNN on galaxies with known stellar mass as target} % : predicting stellar mass
\label{sec:CNN_results_mstar}

\begin{figure}
\centering
\includegraphics[width=\columnwidth]{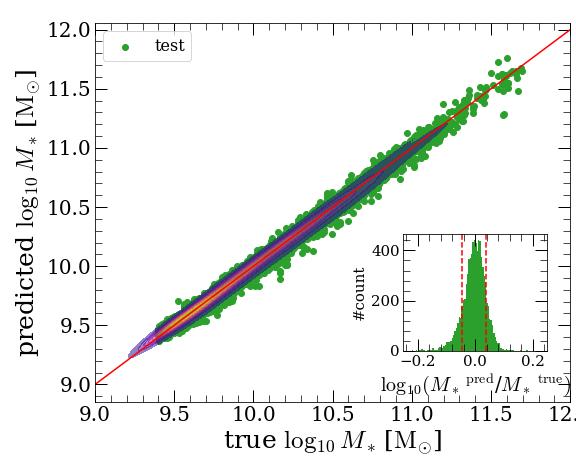}
\includegraphics[width=\columnwidth]{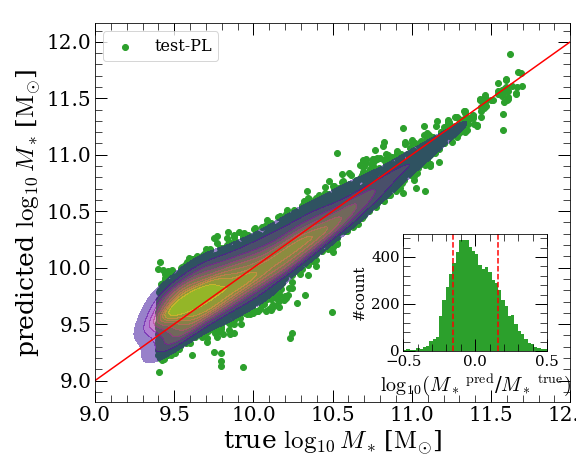}
\caption{Top: Central $M_{*}$ prediction of our CNN-based model ($r$-band image and two velocity maps as input, trained by known central $M_{*}$), as a function of their true value. Bottom: Central $M_{*}$ prediction through power-law fitting, as a function of their true value. In both panels, the red line indicates the prediction equals ground truth, and the green dots are the test set of our samples. The contours indicate the density distribution of the green dots. The histogram in the lower right of each panel shows the distribution of $M_{*}$ prediction over ground truth ratio respectively with the red dashed lines indicating the $1-\sigma$ range. } 
\label{fig:CNN_M*_prediction}
\end{figure}

Using observations, the stellar mass can be estimated through many methods, including SPS analysis through spectrum or SED fitting, multi-component modelling with stellar kinematics, and gravitational lensing measurements. In our CNN model, we train a multi-branch ResNet using $r$-band images and two velocity maps as input, and using known stellar mass $M_*\,(\leqslant R_{\rm hsm})$ (i.e., the stellar mass enclosed within a 3D sphere 
with radius of $R_{\rm hsm}$) as target. This specific setup is to imitate the situation where we only have good knowledge about stellar mass $M_*$ for the training sample. For instance, galaxies in the southern sky may lack high-quality spectroscopic and multi-waveband photometric data, lowering SED estimated galaxy stellar mass reliability.  
The top 
panel of Fig.\,\ref{fig:CNN_M*_prediction} shows the CNN-predicted stellar mass $M_*\,(\leqslant R_{\rm hsm})$ versus the true values for the test set. As shown in Table \ref{tab:CNN_performance_III}, the 1-$\sigma$ scatters in the predicted $\log M_{*}^{\rm pred}/M_{*}^{\rm true} $ values, either within $R_{\rm hsm}$ or within $2R_{\rm hsm}$, are 0.04 dex (i.e., $\sim 10\%$). Under the default image and map conditions, if only $r$-band images, or $g$- and $r$-images combined, or solely two velocity maps are used as input, the uncertainties become slightly larger, of 0.06 dex (i.e., $13\%- 15\%$), 0.05 dex and 0.05 dex, respectively. We note that changing the image resolution also affects the 1-$\sigma$ scatters. Uncertainties increase to 0.06 dex and 0.07 dex (i.e., $15\%-17\%$) when only using $r$-band images with resolutions of 150\,$\times$\,150 and 60\,$\times$\,60 (MaNGA-like), respectively. 

It is interesting to ask, when making a prediction, whether the CNN model simply picks up some general scaling relation between a galaxy's stellar mass and some summary statistics, such as total luminosity, or has it actually used higher-order information encoded in the light and kinematic maps? As a simple test, we fit a power-law relation to the stellar mass and luminosity of galaxies in the training sample and use the best-fit $\log M_* - \log L$ relation to predict stellar masses for the test set luminosities. The relationship is shown at the bottom 
panel of Fig.\,\ref{fig:CNN_M*_prediction}. The scatter in $\log M_{*}^{\rm pred}/M_{*}^{\rm true}$ is 0.16 dex in this case, significantly larger than the scatter of our CNN-based results (see the top panel of Fig.\,\ref{fig:CNN_M*_prediction}). In addition, a decision-tree based regression method, which takes the total luminosity $L$ and velocity dispersion $\sigma_{\rm v}$ as input, also results in a scatter of 0.05 dex (details are presented in Section\,\ref{sec:GBDT_results}). The much smaller uncertainty from our multi-branch ResNet CNN model indicates that the network has actually made better predictions using spatial distributions of light and kinematics. This is similar to the findings of the  \citet{Euclid2023preparationMLHimage}, where they also found that their model predictions of stellar masses improved with the inclusion of image data.
%\citet{Euclid2023preparationMLHimage} also find that prediction of stellar mass improves with image inclusion. 

Since 3D stellar mass can not be obtained in observation without dynamical modelling, we also use 2D cylindrical/projected masses as targets. As shown in Table \ref{tab:CNN_performance_III}, the 1-$\sigma$ scatters in the predicted $\log M_*^{\rm 2D,\,pred}/M_*^{\rm 2D,true}$ values are similar to cases where targets are 3D spherical stellar mass. 

Having obtained stellar masses $M_*$ from CNN based models, we can make further predictions on stellar mass-to-light ratios $M_*/L$ by utilizing $r$-band luminosities $L^{\rm true}$ directly calculated from images (within the same aperture radius) assuming that the luminosity can be well measured. 
%\textcolor{purple}{I personally tend to remove it as well. It is unlikely to receive poor luminosity measurement in optical surveys, though this could happen in MIR and radio observation.}
%\rjlnote{not sure why this is needed - `and assuming that the luminosity can be well measured'}. Specifically, $(M_*/L)^{\rm pred}\equiv M_*^{\rm pred}/L^{\rm true}$. 
In this case, the uncertainties in $\log \big[ (M_*/L) ^{\rm pred} / (M_{*} / L )^{\rm true} \big]$ are then dominated by the uncertainties in CNN-based stellar mass predictions. 
Note that in Section\,\ref{sec:CNN_results_m2l}, we compare between the $M_*/L$ predictions made by CNN models which take $M_*$ as the target (as presented  here) and the predictions made by CNN models that directly take $M_*/L$ as the target. We find that the latter has larger uncertainties than the former. 

\subsection{Training CNN on galaxies with known total mass as target} %: predicting total mass 
\label{sec:CNN_results_mtot}

\begin{figure}
\centering

\includegraphics[width=\columnwidth]{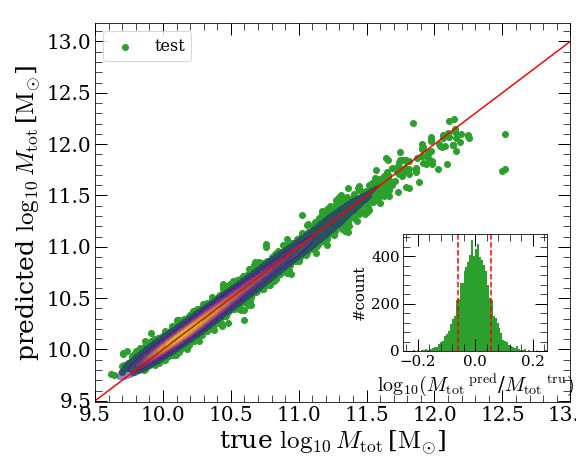}
\caption{Central $M_{\rm tot}$ 
prediction of our CNN-based model ($r$-band image and two velocity maps as input, trained by known central $M_{\rm tot}$), as a function of their true value. The symbols are the same as Fig.\,\ref{fig:CNN_M*_prediction}.}
\label{fig:CNN_mtot_prediciton}
\end{figure}

Galaxies live in dark matter halos. The total mass $M_{\rm tot}$ is a fundamental property of a galaxy. From a dynamical modelling perspective, unlike stellar mass, where the results are degenerate with that of dark matter, the total mass can be more reliably determined through dynamical or lensing modelling approaches (e.g., \citealt{Treu_2010ARA&A_GalStrongLensing, LiHongyu_2016MNRAS_IllustrisJAM, ZhuLing2020_MethodValidation}). This specific CNN model is to mimic the situation where total masses $M_{\rm tot}$ (i.e., the total mass enclosed within a 3D sphere with a radius of $R_{\rm hsm}$) are known and available in the training sample 
, together with their $r$-band images and IFU-like kinematic maps. The predictions on $M_{\rm tot}\, (\leqslant R_{\rm hsm})$ for the test set are given in Fig.\,\ref{fig:CNN_mtot_prediciton}. As shown in Table \ref{tab:CNN_performance_III}, the 1-$\sigma$ uncertainties in $\log M_{\rm tot}^{\rm pred}/M_{\rm tot}^{\rm true}$, as predicted within $R_{\rm hsm}$ and $2R_{\rm hsm}$, are both 0.06 dex ($\sim 15\%$). It is important to realize that taking images and velocity maps together works better than if individual input maps are used alone. Specifically, if only 2D photometric information is used, either taking $r$-band images or taking both $r$- and $g$-band images together, the scatter is 0.07 dex ($\sim 17\%$). If only stellar kinematic maps are used, the scatter is larger at 0.09 dex ($\sim 23\%$).

We note that single- or multi-band images and velocity maps, taking each kind on their own, contain information about the stellar mass and the total mass. However, it is hard for us to answer which kind of map has actually provided more information. This is because the input image and velocity maps have different spatial resolutions. As recorded in \citet{LiHongyu_2016MNRAS_IllustrisJAM}, higher resolution maps result in smaller uncertainties in the estimated dynamical masses of galaxies. Here without carrying out a further resolution test, we cannot make a concrete assessment on this point. However, as we will see in Section\,\ref{sec:GBDT_results}, a decision tree-based method helps us to address this question to the first order, revealing that, by comparison with other galaxy properties, luminosity plays a dominant role in predicting the stellar and total masses.

It is also interesting that, given the same input, the uncertainty in predicted stellar masses is always smaller than in the predicted total masses. This essentially reflects a tighter correlation between a galaxy's stellar mass and its morphology and kinematics, by comparison with the total mass.

\subsection{Training CNN on galaxies with known stellar and total masses as targets} %: predicting stellar and total masses 
\label{sec:CNN_results_2prop}

\begin{figure*}
\includegraphics[width=\columnwidth]{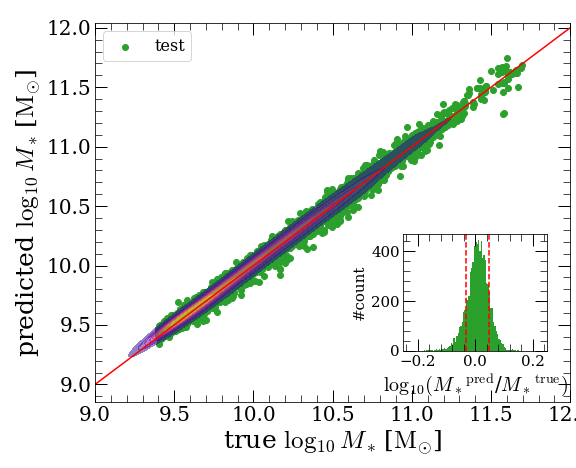}
\includegraphics[width=\columnwidth]{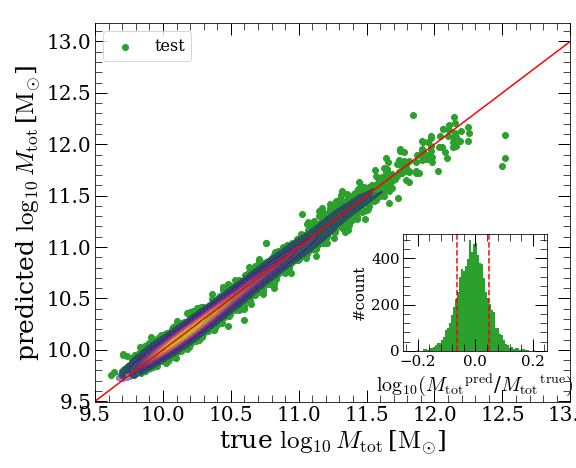}
\includegraphics[width=\columnwidth]{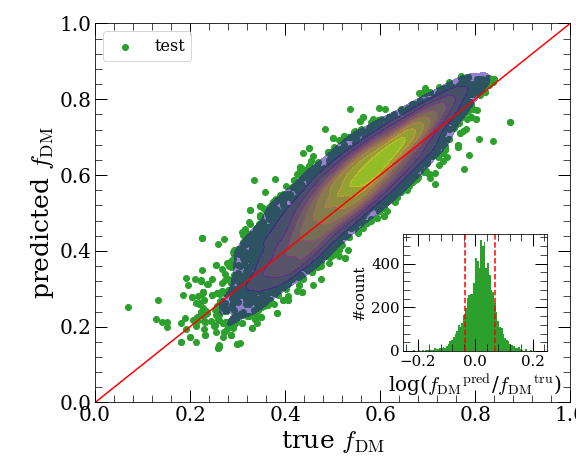}
\includegraphics[width=\columnwidth]{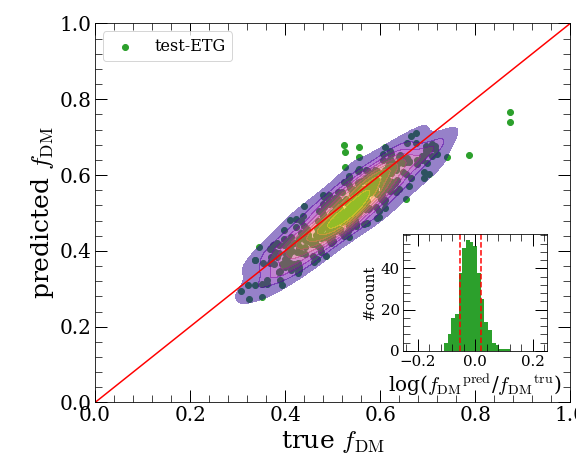}

\caption{Central  $M_{*}$ (top left), $M_{\rm tot}$ (top right), $f_{\rm DM}$ (bottom left), and $f_{\rm DM}$ of selected early-type galaxies (bottom right) predictions of our CNN-based model ($r$-band image and two velocity maps as input, trained by known central $M_{*}$ and $M_{\rm tot}$), as a function of their true values. Here $f_{\rm DM}$ is the dark matter fraction, calculated by $f_{\rm DM}=1-M_*/M_{\rm tot}$. The symbols are the same as Fig.\,\ref{fig:CNN_M*_prediction}. We note that the bottom panels present $f_{\rm DM}$ in linear scales and therefore the distributions appear wider than those for logarithmic masses in the upper panels. }
\label{fig:CNN_2prop_prediction}
\end{figure*}

From observations, we can obtain both the stellar mass $M_*$ and the total mass $M_{\rm tot}$ for a galaxy, either through individual estimates as mentioned above,  or jointly through multiple dynamical tracers. Alternatively, such information can be acquired by using galaxy formation and evolution models. This specific CNN model is to imitate this situation, where both quantities are available in the training sample. The top panels in Fig.\,\ref{fig:CNN_2prop_prediction} show the performance of the trained multi-branch ResNet in simultaneously predicting 3D spherical $M_{*}(\leqslant R_{\rm hsm})$ and $M_{\rm tot}(\leqslant R_{\rm hsm})$ for the test set.  
The 1-$\sigma$ uncertainties in $\log M_{*}^{\rm pred}/M_{*}^{\rm true}$ and $\log M_{\rm tot}^{\rm pred}/M_{\rm tot}^{\rm true}$ are 0.04 dex and 0.06 dex, respectively. However, some biases of $\pm$0.01 dex are noticed in this model. It is interesting to note that employing both quantities at the same time as targets in the model does NOT increase prediction accuracy significantly, by comparison with only utilizing one type of target at a time (see Sections \ref{sec:CNN_results_mstar} and \ref{sec:CNN_results_mtot}). 

When reliably quantified observational systems are used as training data, the overall model performance cannot exceed the accuracy on the training data. However, it is interesting to make comparisons between model predictions and conventional estimates over a generalized statistical population. As the latter suffer from various systematics that vary differently from galaxy to galaxy (as already discussed in Section\,\ref{sec:introduction} ), here, we compare the mass estimation accuracy between CNN-based models and JAM-based models, given that the same kinds of input information are used, i.e., single-band images and IFU-like kinematic maps. \citealt{LiHongyu_2016MNRAS_IllustrisJAM} evaluated the performance of JAM using galaxies from the Illustris cosmological simulations\,\citep{nelson2015Illustris} 
, assuming MaNGA-like image and IFU observational conditions. The typical scatter of JAM-based total masses is about 11-16\%, i.e., the scatter on total mass estimates from the two approaches is comparable. While CNN models in general predict stellar masses with higher accuracies ($\sim$ 10\%) than total masses, JAM-based predictions are the opposite. The stellar masses predicted by JAM modelling often suffer from much larger uncertainties with $\sim$30\% scatter due to model degeneracies between the stellar component and dark matter. \citet{Zhu2023_MaNGADynPopI} 
adopted six different composite models describing dark matter and baryon distributions and fitted the models to the MaNGA galaxies for which reliable measurements for IFU kinematics are available. The mean standard deviation in predicted stellar masses across the six models over the full galaxy sample is 0.19 dex ($\sim 50\%$). We treat this scatter between different models as possible model uncertainties due to unknown degeneracies and hidden systematics. By comparison, our CNN models predict stellar masses with an uncertainty of 0.07 dex ($15\sim17\%$) for the entire population. Our superiority on CNN stellar mass accuracy might be as a result of the complexity of the neural network which encodes knowledge of the stellar masses for the training sample. We note that, however, a fair comparison between conventional methods and our CNN methods should be made with the same data sample in order to draw more concrete conclusions.

We calculate dark matter fractions $f_{\rm DM}\,(\leqslant R_{\rm hsm})$ 
based on our CNN model predictions for $M_*$ and $M_{\rm tot}$, and compare the fractions with their true values. Here we assume that the dark matter fraction is simply given by $f_{\rm DM}=1-f_*$, where $f_* \equiv M_*/M_{\rm tot}$. As expected, such an approximation on the dark matter faction would be an overestimate for galaxies with a significant amount of central gas. This is manifested by $\log f_{\rm dm}^{\rm pred}/f_{\rm dm}^{\rm true} \sim 0.02 \pm 0.05$\,dex for the overall sample. When we select only early-type galaxies (as defined in \citealt{2020MNRAS.491.5188W}) and carry out the same estimate (but without re-training the CNN model), as can be seen in the bottom panels of Fig.\,\ref{fig:CNN_2prop_prediction}, the scatter becomes markedly narrower by comparison with the overall sample. In this latter case, however, the dark matter fraction for this sub-sample is underestimated by 0.02 dex. This is because the model predicted $M_{\rm tot}$ was underestimated by 0.02 dex for these galaxies.

\subsection{Training CNN on galaxies with known stellar mass to light ratio as target} %: predicting stellar mass to light ratio
\label{sec:CNN_results_m2l}
\begin{figure*}
\centering

\includegraphics[width=\columnwidth]{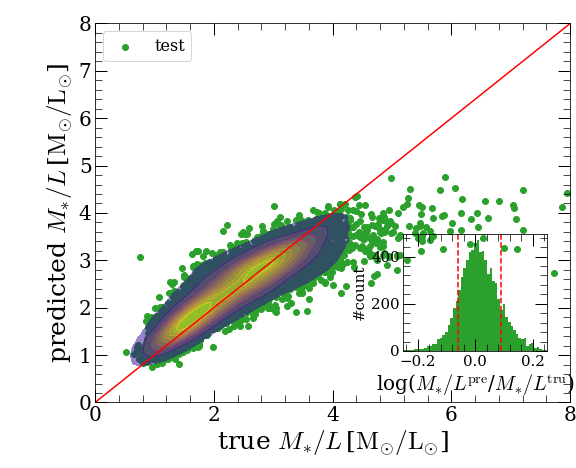}
\includegraphics[width=\columnwidth]{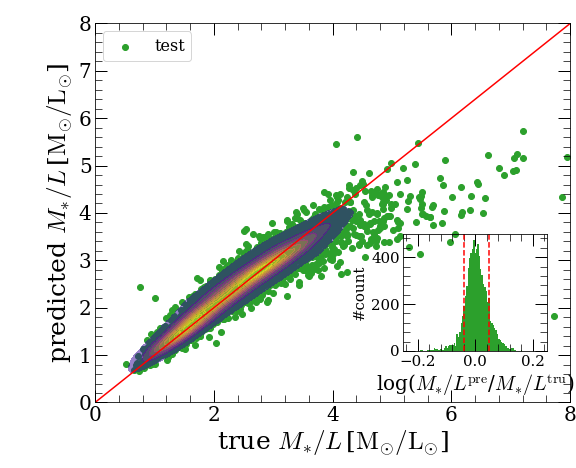}
\caption{Central $M_*/L$ prediction ($r$-band image and two velocity maps as input, trained by known central $M_*/L$; left) and central $M_*/L$ prediction ($g$- and $r$-band images and two velocity maps as input, trained by known central $M_*/L$; right) of our CNN-based model, as a function their true value respectively. The symbols are the same as Fig.\,\ref{fig:CNN_M*_prediction}.}
\label{fig:CNN_m2l_prediction}
\end{figure*}

A recent study by \citet{dobbels2019} using a machine learning approach showed that morphological information from galaxy $g$-band images can noticeably improve the determination of galaxies' $M_*/L$s, by comparison with those obtained from only one or two colours. Specifically, they used convolutional neural networks to learn key morphological features in the $g$-band images, which were then fed into a gradient boosting model to predict the stellar mass-to-light ratios $M_*/L$. This two-step algorithm was trained on a sample of more than 80,000 galaxies from the GALEX-SDSS-WISE Legacy Catalog version 2 \citep[GSWLC;][]{salim2016gswlc,salim2018gswlc}. The ground-truth $M_*/L$s were determined by global spectral energy distribution fitting. The uncertainty in $M_*/L$s for the observed galaxy sample was $\sim$\,0.15 dex. Their investigation has already shed light on a feasible way to use machine learning to find connections between $M_*/L$ and galaxy 2D mass and light distributions.

In this work, we train a CNN model to directly predict $M_*/L$s in the case where both $r$-band images and IFU-like kinematic maps are available. In this case, $M_*/L$ is defined as the ratio between the projected 2D stellar mass and $r$-band luminosity within a given radius. Our results are given in Fig.\,\ref{fig:CNN_m2l_prediction}. As can be seen, the scatter in $\log \big[ (M_*/L)^{\rm pred} / (M_*/L)^{\rm true} \big]$ is about 0.07 dex  ($\sim 15\%$, for within both $R_{\rm hsm}$ and $2R_{\rm hsm}$). By comparison, the scatter is about 0.1 dex if we only take $r$-band images, or only take two velocity maps, as input.

The scatters in $\log \big[ (M_*/L)^{\rm pred} / (M_*/L)^{\rm true} \big] $ we obtain are generally much larger than those estimated via $M_*$ in the previous sections (see Section\,\ref{sec:CNN_results_mstar} and \ref{sec:CNN_results_2prop}). This indicates that, under the input conditions used and with the same network complexity, using $M_*$ as the CNN model target yields better predictions on $M_*/L$ than directly using $M_*/L$ as the target.

An additional investigation is to add images in another band such that colour information is also available to the model network. As \citet{BellJong01Color-M2L} reveal, galaxy $M_*/L$ strongly correlates with the galaxy's colour. Indeed, as is shown in Section\,\ref{sec:GBDT_results_ratio}, our GBDT results also reveal that a galaxy's colour is a key contributing factor in making a correct prediction for $M_*/L$. 
We took a galaxy's $g$-band image as an additional input to our multi-branch ResNet. When both $g$- and $r$-band images are used, the scatter in $\log \big[ (M_*/L)^{\rm pred} / (M_*/L)^{\rm true} \big]$ is reduced to 0.05 dex. If the colour information is further combined with two velocity maps, the scatter then reduces to 0.04 dex. In both cases, the scatters are significantly smaller than 0.1 dex when only $r$-band images, or only kinematic maps, were used, or smaller than 0.07 dex when $r$-band images and kinematic maps combined were used.

\section{Light-GBM results}
\label{sec:GBDT_results}

Having demonstrated in the previous section the abilities of our CNN models to predict galaxy masses, we apply a Gradient Boosting Decision Tree (GBDT) method to investigate the driving factors in making successful predictions based on spatially resolved light and kinematic distributions. To do so, we take a gradient boosting machine (Light-GBM), and train a model to compute feature importance. We use `gain' importance - the total gains of conditions in the model which use a feature\footnote{\url{https://lightgbm.readthedocs.io/en/latest/pythonapi/lightgbm.Booster.html\#lightgbm.Booster.feature_importance}}. In case any pair of linearly-correlated features/targets might bias the feature importance evaluation, we also compute linear correlation coefficients between features and targets (Section~\ref{sec:correlation matrix}). Fig.\,\ref{fig:paper_structure} summarises our GBDT training workflow, starting from calculating input summary statistics (listed in Table~\ref{tab:GBDT_input}) from images and maps, to model training and final property predictions. Table \ref{tab:GBDT_loss} shows the mean and standard deviation (uncertainty) of the predicted properties from different GBDT models. Detailed results are presented below.

\begin{table*}
    \caption{Results of GBDT model}
    \centering
    
    \begin{tabular}{c|c|c|c|c}
      \hline Section & Input & Predictions & Mean $\pm$ 1-$\sigma$ (dex) \\
      \hline  \ref{sec:GBDT_results_mass}& default & $M_*$ &0.00 $\pm$  0.05  \\
      \hline  \ref{sec:GBDT_results_mass}& default w/o Mag & $M_*$ & 0.00 $\pm$  0.08  \\
      \hline  \ref{sec:GBDT_results_mass}& default &  $M_{\rm DM}$ &0.00 $\pm$  0.13  \\
      \hline  \ref{sec:GBDT_results_mass}& default w/o Mag& $M_{\rm DM}$ & 0.00 $\pm$  0.15  \\
      \hline  \ref{sec:GBDT_results_mass}& default & $M_{\rm tot}$ &0.00 $\pm$  0.08  \\
      \hline  \ref{sec:GBDT_results_mass}& default w/o Mag& $M_{\rm tot}$ & 0.00 $\pm$  0.11  \\
      \hline  \ref{sec:GBDT_results_ratio}& default &  $f_*$ &0.01 $\pm$  0.08  \\
      \hline  \ref{sec:GBDT_results_ratio}& default & $f_{\rm DM}$ & 0.00 $\pm$  0.06  \\
      \hline  \ref{sec:GBDT_results_ratio}& default & $M_*/L$ & 0.00 $\pm$  0.06  \\
     
        \hline
    \end{tabular}
    \caption{Performance of GBDT methods. The fourth column indicates the mean and standard deviation of the logarithmic ratio between the predicted and the true values for quantities given in the third column. All properties are evaluated within a radius of $R_{\rm hsm}$ from the galaxy centre. }
    \label{tab:GBDT_loss}
\end{table*}

\subsection{Linear correlation between features and targets}
\label{sec:correlation matrix}

\begin{figure*}
\centering
\includegraphics[width=\textwidth]{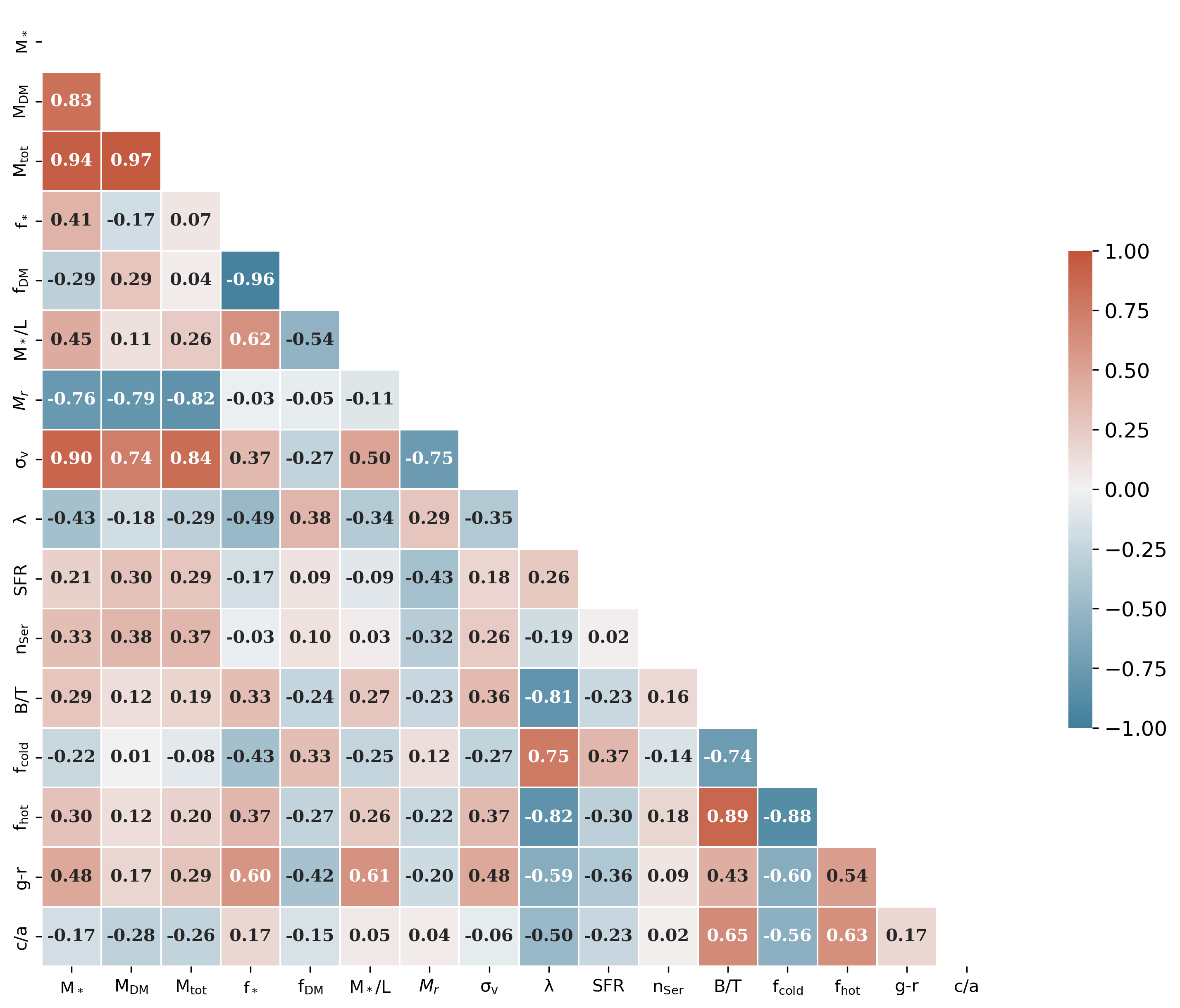}
\caption{The correlation matrix of features and targets we used for LightGBM-based model training (Section~\ref{sec:GBDT}) and evaluation (Section~\ref{sec:GBDT_results}). A variable-pair on the diagram would have a higher possibility of being linearly correlated if the absolute value of its correlation coefficient is closer to 1.}
\label{fig:correlation_matrix}
\end{figure*}

Before assessing non-linear feature importance using the Light-GBM algorithms when predicting masses (Section~\ref{sec:GBDT_results_mass}) and ratios/fractions (Section~\ref{sec:GBDT_results_ratio}), we first evaluate possible linear correlations between pairs of features or targets. Similarly to \citet{vonMarttens2022}, we compute the Pearson correlation coefficient ({\it R}). 
In statistics, computing {\it R} (ranging from -1 to 1) between two variables in a sample can help assess if there exists a linear correlation between them. The two variables are more likely to be positively linearly correlated if the {\it R} is close to 1, or negatively correlated when close to -1. The results of such computations can be seen as a correlation matrix in Fig.\,\ref{fig:correlation_matrix}.  

Looking at the input features, we find that 
pairs of features share reasonably strong linear (anti-) correlations (|{\it R}| > 0.75) including $\sigma_{\rm v}$ - luminosity, $\lambda$ - $B/T$, $\lambda$ - $f_{\rm cold}$, $\lambda$ - $f_{\rm hot}$, $B/T$ - $f_{\rm hot}$ and $f_{\rm cold}$ - $f_{\rm hot}$. 
The {\it R} value between $B/T$ and $f_{\rm cold}$ is also close to 0.75 (${\it R}$ = 0.74). 
In other words, $B/T$, $\lambda$, $f_{\rm cold}$ and $f_{\rm hot}$ are generally inter-correlated with each other. 
Physically, a fast-rotating galaxy is expected to have a higher stellar spin $\lambda$ as well as a higher cold-orbit fraction $f_{\rm cold}$.  The galaxy would also have a smaller kinematic $B/T$ and a smaller hot-orbit fraction $f_{\rm hot}$. 
The strong correlation between  $\sigma_{\rm v}$ and luminosity essentially reflects the Tully-Fisher relationship and the Faber-Jackson relationship obeyed by the simulated galaxies (see \citealt{LuShengdong_2020MNRAS_TNG_FP} for a detailed discussion on the fundamental plane properties of TNG100 galaxies). 
As can be seen in Fig.\,\ref{fig:correlation_matrix}, other features do not show strong evidence of linear correlations. 

When we look at targets and features together, we find luminosity and $\sigma_{v}$ are generally strongly correlated (mostly with their |{\it R}| > 0.75, though {\it R} for $\sigma_{v}$-$M_{\rm DM}$ equals 0.74) with the masses ($M_{*}$, $M_{\rm DM}$ and $M_{\rm tot}$), indicating they play a dominant role in predicting these mass values. However, as stated in \citet{vonMarttens2022}, 
{\it R} can only indicate a linear relationships between two variables. The evaluation of the non-linear relationships requires other methods, which are discussed in the following section.

\subsection{Dependencies of the total, stellar, and dark matter masses}
\label{sec:GBDT_results_mass}

\begin{figure*}
\centering
\includegraphics[width=0.95\textwidth]{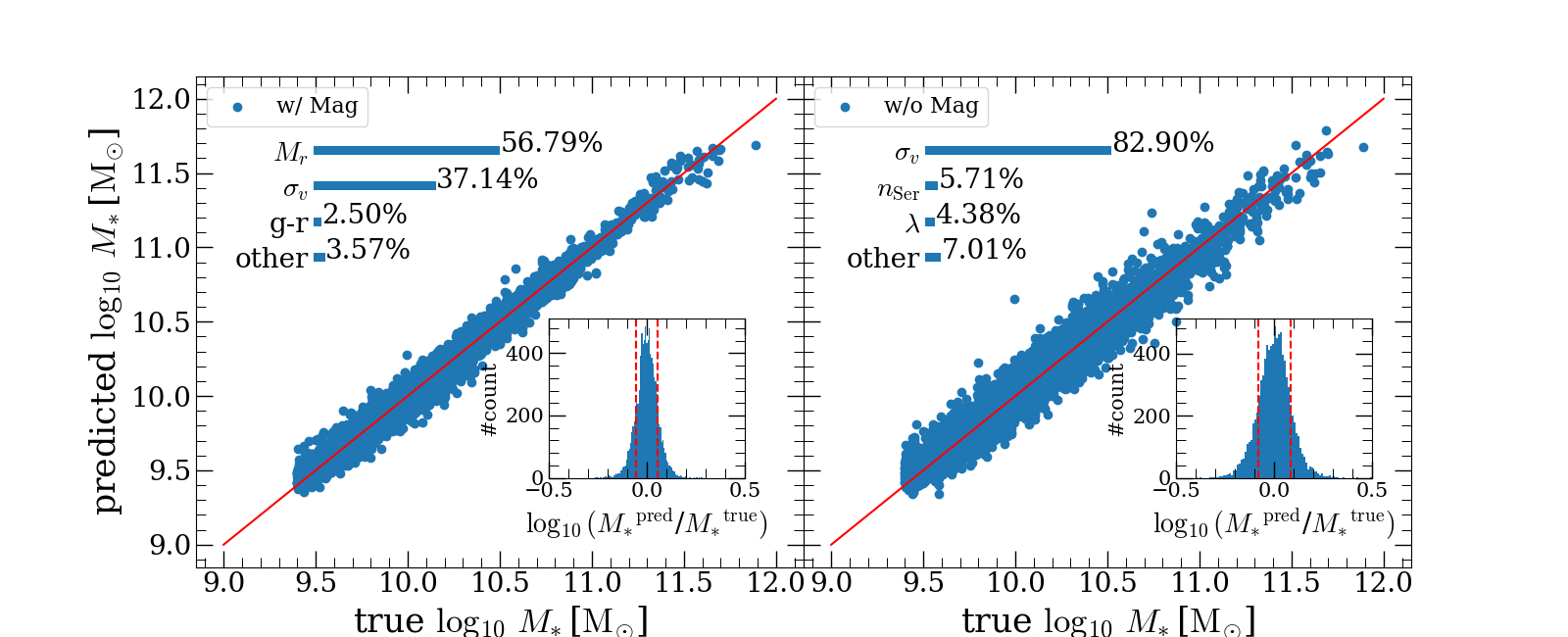} \\
\includegraphics[width=0.95\textwidth]{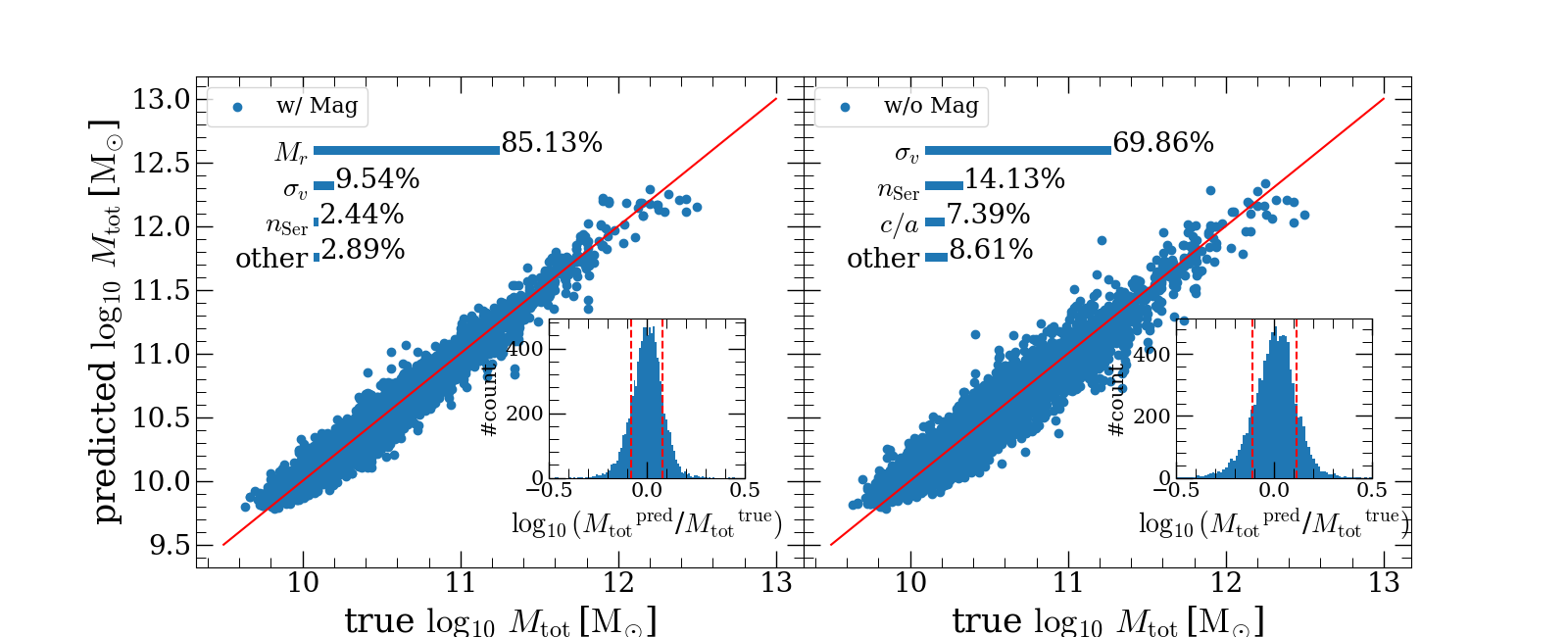}\\
\includegraphics[width=0.95\textwidth]{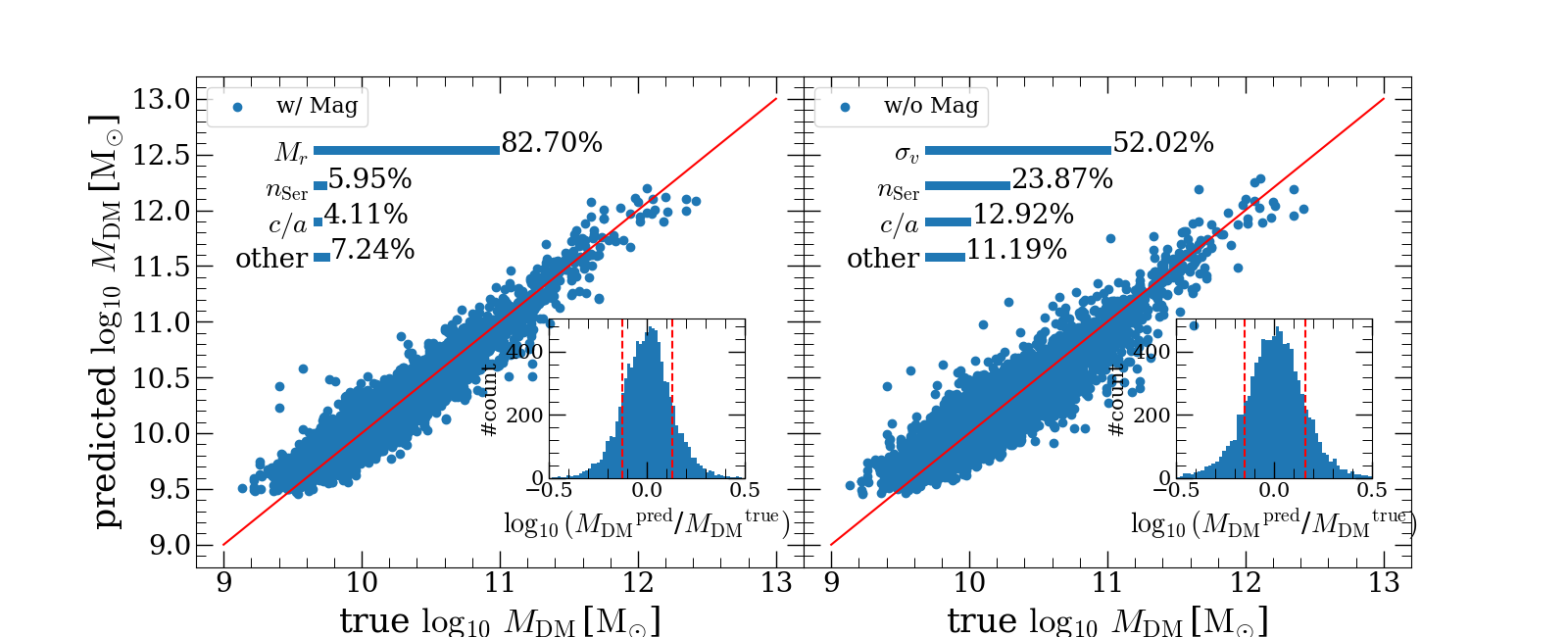}
\caption{From top to bottom: central $M_*$, $M_{\rm tot}$, and $M_{\rm DM}$ predictions (trained by known $M_*$, $M_{\rm tot}$ and $M_{\rm DM}$ respectively), as function of their true value. In each row, the left panel has taken $r$-band magnitude as one of the input features, while the right panel has not. In all 6 panels, our GBDT models are trained independently. The red line indicates that the prediction equals ground truth, and the blue dots are the samples of our test set. The histogram at the lower right of each panel shows the distribution of mass prediction over the ground truth ratio respectively, with the red dashed line indicating 1$\rm \sigma$ range. The bar graph at the upper left of each panel shows the important features and their contributions to the GBDT predictions.}

\label{fig:GBDT_results_mass}
\end{figure*}

We train our GBDT model to predict separately the 3D spherical $M_*$, $M_{\rm DM}$, $M_{\rm tot}$ values within a radius  
of $R_{\rm hsm}$ from the galaxy centre. The outcome is shown in the left column of Fig.\,\ref{fig:GBDT_results_mass}. The scatter for the stellar and total masses are 0.05 dex and 0.08 dex, respectively. For dark matter mass, the uncertainty is about 0.13 dex. It is interesting to note that, among all the properties investigated, galaxy luminosity (magnitude) contributes the most to all three mass predictions. Having trained the GBDT model without using magnitude as input, the results are presented in the right column of Fig.\,\ref{fig:GBDT_results_mass}. As can be seen, the second most significant feature that contributes to mass predictions is velocity dispersion.

\subsection{\texorpdfstring{Dependencies of ratios and fractions: $M^*/L$, $f_{*}$ and $f_{\rm DM}$}{Dependencies of ratios and fractions: M*/L, f* and fDM}}
\label{sec:GBDT_results_ratio}
\begin{figure}
\centering
\includegraphics[width=\columnwidth]{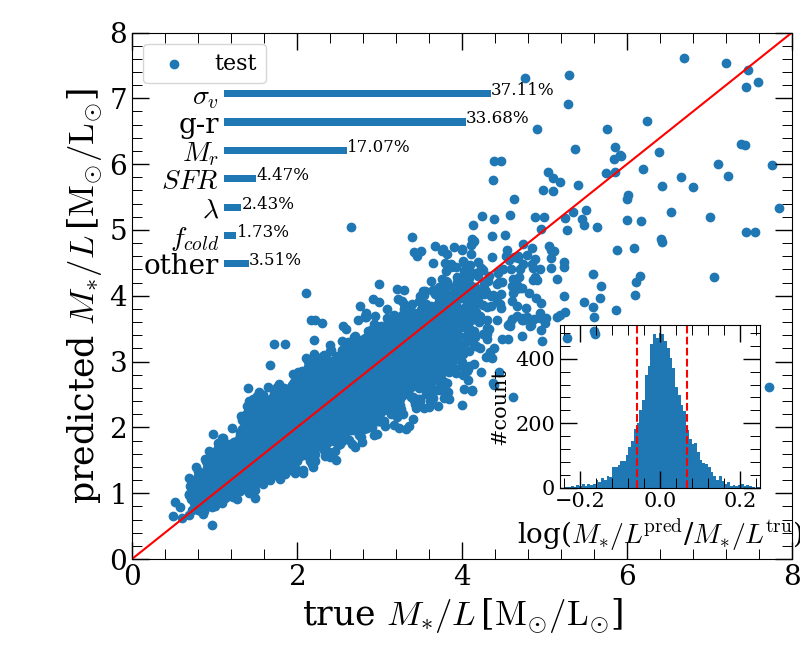} \\
\includegraphics[width=\columnwidth]{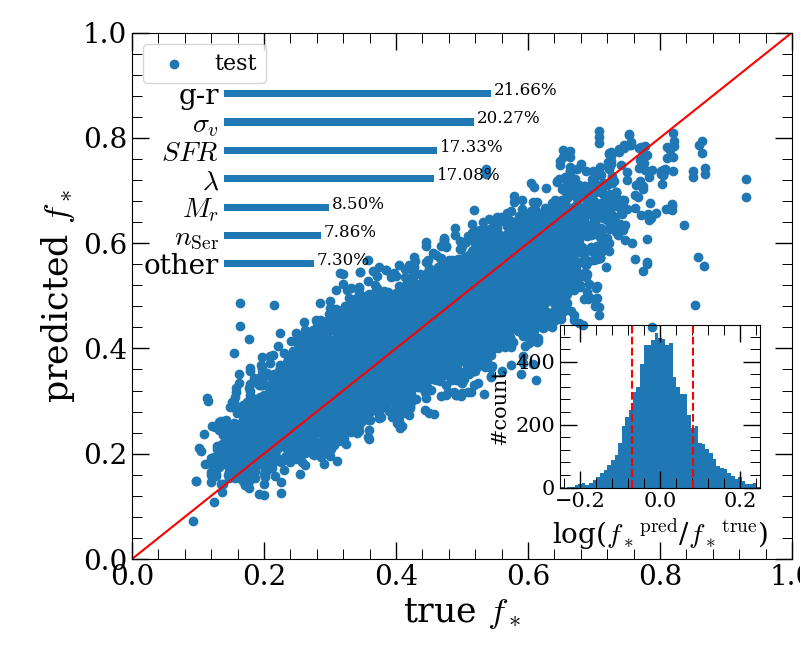}\\
\includegraphics[width=\columnwidth]{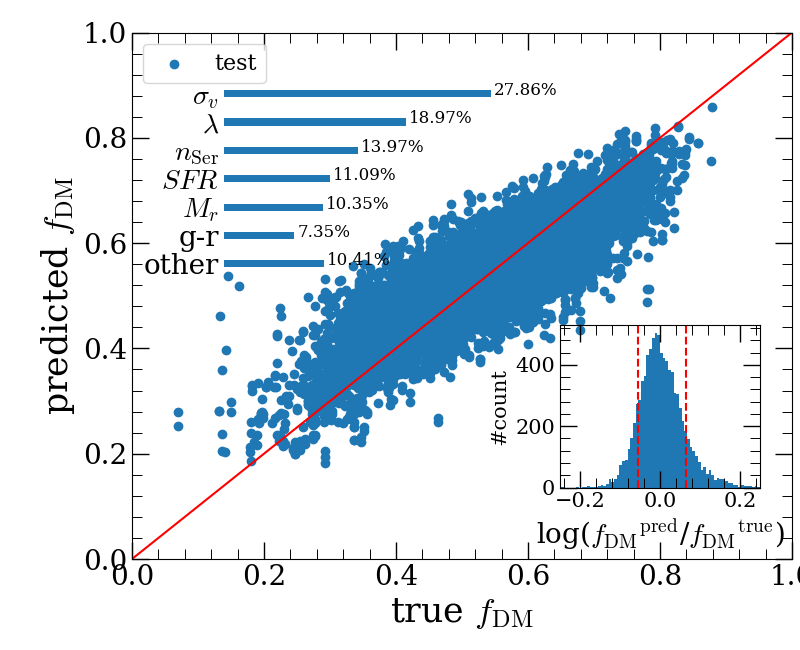}
\caption{From top to bottom: central $M_*/L$, $f_{*}$, and $f_{\rm DM}$ predictions (with $r$-band magnitude and other summary statistics as input, trained by known $M_*/L$, $f_{*}$, and $f_{\rm DM}$ respectively), as a function of their true value. The symbols are same as Fig\ref{fig:GBDT_results_mass}.}
\label{fig:GBDT_results_ratio}
\end{figure}

We train GBDT models to predict $M_*/L$, $f_{*}$ and $f_{\rm DM}$ within a radius of $R_{\rm hsm}$ from the galaxy centre. Fig.\,\ref{fig:GBDT_results_ratio} shows the results from the best-trained models on the test set. For the stellar and dark matter mass fractions $f_{*}$ and $f_{\rm DM}$, the uncertainties are 0.08 dex and 0.06 dex, and the uncertainty of the mass-to-light ratio $M_*/L$ is 0.06 dex. It is important to note that CNN model predictions are much more accurate than GBDT predictions as produced in this study. This is unsurprising because the former provides information on the spatial distribution of galaxy properties, while the latter only takes low-order summary statistics into account.   

In our models, feature importance shows that the contributions are complicated. Unlike mass predictions, there is no dominant feature in the ratio predictions. 
Generally, velocity dispersion contributes the most, $\sim 37 \%$ in predicting $M_*/L$ and $\sim 28 \%$ in $f_{\rm DM}$, with the other features having smaller contributions. 
For $f_*$ and $M_*/L$, velocity dispersion and a galaxy's colour are the top two contributing features. 
The stellar spin parameter $\lambda$, which reflects the dynamical status of a galaxy, is the second most important feature in predicting $f_{\rm DM}$. 

\section{Discussion and conclusions}
\label{sec:conclusion}

In this study, we use a general sample of galaxies from the TNG100 simulation to investigate the ability of our CNN-based models to predict the central (i.e., within $1-2R_{\rm hsm}$) stellar mass, total mass, stellar mass-to-light ratio, and to estimate the dark matter fraction. Specifically, we take galaxy images, spatially-resolved mean velocity and velocity dispersion maps as input to our multi-branch ResNet CNN models (see Section\,\ref{sec:CNN_input} for detailed data input and target generation; the detailed method is given in Section\,\ref{sec:CNN}). In particular, the IFU-like kinematic maps have spatial resolution typical of the MaNGA galaxy sample, and cover a square region of $[-3R_{\rm hsm},\,3R_{\rm hsm}]^2$ around the galaxy centre. The CNN-based models, with the help of the training data set, can in general break the degeneracy between the baryon and dark matter distributions and make reliable mass predictions. 
In order to understand which (global) features contribute the most to our predictions, we utilize a gradient boosting machine Light-GBM, which takes global galaxy properties as input, including luminosity, colour, SFR, Sersic index, axis ratio, stellar velocity dispersion, spin parameter, kinetic B/T, and orbital fractions (see Section\,\ref{sec:GBDT_input} for detailed data input and target generation; the detailed method is given in Section\,\ref{sec:GBDT}). 

Our main results are listed below.
\begin{enumerate}
    \item Our multi-branch ResNet CNN models can predict (central) stellar and total masses of galaxies with 1-$\sigma$ uncertainties of 0.04 and 0.06 dex, respectively, when taking $r$-band images and two velocity maps as input. Under such conditions, the prediction for $M_*/L$ has an uncertainty of 0.07 dex. However, when combined with galaxy colour information, e.g., taking both  $g$- and $r$-band images together with kinematic maps as input, the uncertainty decreases to 0.04 dex (for more details see Table\,\ref{tab:CNN_performance_III} in Section\,\ref{sec:CNN_results}.).  
    \item Given the default input to the GBDT models, the stellar and total masses of galaxies can be reproduced with uncertainties of 0.05 and 0.08 dex, respectively. The predicted dark matter mass uncertainty is somehow larger at 0.13 dex. The uncertainties on the central stellar ($f_*$) and dark matter ($f_{\rm DM}$) fractions are 0.08 dex and 0.06 dex, respectively; while that for $M_*/L$ is 0.06 dex (for more details see Table\,\ref{tab:GBDT_loss} in Section\,\ref{sec:GBDT_results}).
    \item We find from our GBDT models that galaxy luminosity is the dominating feature (contributed > 50\%) in predicting all masses in the central $1-2R_{\rm hsm}$ regions (see the left column of Fig.\,\ref{fig:GBDT_results_mass}). When galaxy luminosity is not considered as an input of our GBDT models, the dominating feature  is velocity dispersion. In the case of $f_*$, $f_{\rm DM}$ and $M_*/L$ predictions, we do not observe the existence of a dominating feature (see Fig.\,\ref{fig:GBDT_results_ratio}). Velocity dispersion and galaxy's colour are the top two contributing features when predicting $f_*$ and $M_*/L$. Regarding $f_{\rm DM}$ prediction, we find velocity dispersion contributed the most. At the same time, stellar spin parameter $\lambda$ should also be valued, given it ranked as the second most important feature on the diagram (the bottom panel of Fig.\,\ref{fig:GBDT_results_ratio}).

\end{enumerate}

We note that a galaxy's luminosity is the dominant feature in predicting all masses. In particular, the correlation between luminosity and stellar mass is even tighter than that with the total mass. This can be seen from both the CNN and GBDT model results such that predictions on the stellar mass always have smaller uncertainties than those on the total mass. The tighter connection between the luminosity and the stellar mass can be understood as a consequence of a straightforward conversion through the stellar mass-to-light ratio, which is governed by stellar evolution physics and typically 
spans less than one order of magnitude in value. The connection to the total mass can be understood as a consequence of the fact that observed galaxies obey a certain fundamental plane relation, and, through successful simulation calibration, the galaxy sample has thus implicitly reinforced a correlation between the luminosity and the total mass, which additionally is further subtly influenced by the dynamical interplay between baryons and dark matter. 

Predictions on fractional masses ($f_*$ and $f_{\rm DM}$) and on the mass-to-light ratios ($M_*/L$) show a significant dependence on stellar velocity dispersion (as the leading feature), which reflects the fact that the detailed balance between baryons and dark matter and among different stellar populations, to the first order, have a mass dependence -- a consequence of the hierarchical galaxy assembly history. We also found colour significantly contributed to the stellar predictions ($f_*$ and $M_*/L$) and the stellar spin parameter (which reflects the dynamical nature of a galaxy) to the central $f_{\rm DM}$ prediction, essentially reflect the different physical mechanisms that shape the target properties of the baryonic and dark matter components.

The investigation in this study is in a way reassuring that galaxy images and stellar kinematic maps can provide sufficient information to disentangle the individual dynamical effects from baryons and dark matter. 
However, one must note that any training sample-based CNN in principle cannot reach an accuracy that exceeds that for the training set itself. It is not only hard to obtain an observational galaxy sample with unbiasedly estimated properties, but also impossible to reach accurate predictions for a given sample of observed galaxies by directly applying models that are trained using simulation data, and without taking observational effects and selection rules into account. In addition, data uncertainties and uncertainties in the IMF and galaxy formation physics may cause biases and systematics when making predictions. One potential way to help bridge the gap between simulations and observations may be to test models trained on one simulation with another simulation where different galaxy formation physics have been implemented, though this is yet to be considered in any detail. In this regard, great efforts are still required to find machine learning models that can unbiasedly estimate matter composition for observed galaxies, especially for those machine learning methods using image data as input.

\section*{Acknowledgements}
The authors thank the anonymous reviewers for their useful comments. 
The authors acknowledge the Tsinghua Astrophysics High-Performance Computing platforms at Tsinghua University for providing computational and data storage resources that were utilized in producing the research results reported within this paper. JNC wishes to thank Shude Mao, Boyi Ding, Haoxiong Liu, Zechang Sun, Ce Sui, Kangning Diao, Tao Jing, and Xiaosheng Zhao for various fruitful discussions, and appreciates the machine learning sessions in the Department of Astronomy (DoA) at Tsinghua. HT gratefully acknowledges support from the Shuimu Tsinghua Scholar Programme of Tsinghua University, the China Postdoctoral Science Foundation fellowship 2022M721875, and long-lasting support from various machine learning groups notably the University of Manchester Jodrell Bank Centre for Astrophysics, the TAGLAB research group, and the DoA at Tsinghua. 

%%%%%%%%%%%%%%%%%%%%%%%%%%%%%%%%%%%%%%%%%%%%%%%%%%
\section*{Data Availability}

The data underlying this article will be shared after a reasonable request to the corresponding author.
% The inclusion of a Data Availability Statement is a requirement for articles published in MNRAS. Data Availability Statements provide a standardised format for readers to understand the availability of data underlying the research results described in the article. The statement may refer to original data generated in the course of the study or to third-party data analysed in the article. The statement should describe and provide means of access, where possible, by linking to the data or providing the required accession numbers for the relevant databases or DOIs.



\bibliographystyle{mnras}
% \bibliography{example} % if your bibtex file is called example.bib
\bibliography{reference}

%\begin{thebibliography}{99}
%\end{thebibliography}

%%%%%%%%%%%%%%%%%%%%%%%%%%%%%%%%%%%%%%%%%%%%%%%%%%

%%%%%%%%%%%%%%%%% APPENDICES %%%%%%%%%%%%%%%%%%%%%

\appendix
% \section{label distribution}
% \begin{figure}
%     \centering
%     \includegraphics[width=1\columnwidth]{Plots/dataset/mstar.pdf}
%     \caption{ histgram on $M_*$}
%     \label{fig:Mstar_distribution}
% \end{figure}

% \begin{figure}
%     \centering
%     \includegraphics[width=1\columnwidth]{Plots/dataset/fdm.pdf}
%     \caption{ histgram on $f_{\rm dm}$}
%     \label{fig:fdm_distribution}
% \end{figure}

% \begin{figure}
%     \centering
%     \includegraphics[width=1\columnwidth]{Plots/dataset/m2l.pdf}
%     \caption{ histgram on $M_*/L$}
%     \label{fig:m2l_distribution}
% \end{figure}

% \begin{figure}
%     \centering
%     \includegraphics[width=1\columnwidth]{Plots/dataset/m200.png}
%     \caption{ histgram on $M_{halo}$}
%     \label{fig:m200_distribution}
% \end{figure}
% \section{Some extra material}

% If you want to present additional material which would interrupt the flow of the main paper,
% it can be placed in an Appendix which appears after the list of references.

%%%%%%%%%%%%%%%%%%%%%%%%%%%%%%%%%%%%%%%%%%%%%%%%%%

% Don't change these lines
\bsp	% typesetting comment
\label{lastpage}
\end{document}